\documentclass{svjour3}                     
\smartqed  
\usepackage[dvipdfm]{graphicx}
\graphicspath{{fig/}}
\usepackage{amsmath}
\begin{document}

\title{Weakly nonlinear analysis of two dimensional sheared granular flow}
\author{Kuniyasu Saitoh \and Hisao Hayakawa}
\institute{
Kuniyasu Saitoh
\at
Yukawa Institute for Theoretical Physics, Kyoto University, Sakyoku, Kyoto 606-8502, Japan\\
\emph{Present address:}
Faculty of Engineering Technology, University of Twente, 7500 AE Enschede, Netherlands\\
\email{k.saitoh@utwente.nl}
\and
Hisao Hayakawa
\at
Yukawa Institute for Theoretical Physics, Kyoto University, Sakyoku, Kyoto 606-8502, Japan\\
\email{hisao@yukawa.kyoto-u.ac.jp}
}

\date{Received: date / Accepted: date}

\maketitle

\begin{abstract}
Weakly nonlinear analysis of a two dimensional sheared granular flow is carried out
under the Lees-Edwards boundary condition.
We derive the time dependent Ginzburg-Landau (TDGL) equation of a disturbance amplitude
starting from a set of granular hydrodynamic equations
and discuss the bifurcation of the steady amplitude in the hydrodynamic limit.
\end{abstract}

\section{Introduction}\label{intro}

To control flows of granular particles is important in science and industry \cite{luding1,luding2,bri,gold}.
However, the properties of granular flow have not been well understood yet, because they behave as unusual fluids \cite{jeager}.
This unusual nature is mainly caused by inelastic collisions between granular particles.
Indeed, there is no equilibrium state in granular materials because of  inelastic collisions between grains, 
which suggests that granular materials are an appropriate target of nonequilibrium statistical mechanics \cite{chong}.

Although there are many studies of granular flows on inclined planes \cite{fort,poul},
the existence of gravity and the role of bottom boundary make the problem complicated.
On the other hand, the granular flow under a plane shear is the simplest and an appropriate situation for theoretical analysis.
Therefore, granular flows under a plane shear have been studied from many aspects such as the application of kinetic theory\cite{sela,santos},
shear band formation in moderate dense granular systems \cite{tan,saitoh},
long-time tail and long-range correlation function \cite{kumaran1,kumaran2,kumaran3,orpe1,orpe2,rycroft,lutsko0,otsuki0,otsuki1,otsuki2},
pattern formation of dense flow \cite{louge1,louge2,louge3,louge4,khain1,khain2},
determination of constitutive equation for dense flow \cite{midi,cruz,hatano1},
as well as jamming transition \cite{hecke,hatano2,hatano3,otsuki3,otsuki4,otsuki5}.

In this paper, we focus on the shear band formation in moderate dense granular gases
observed in the discrete element method (DEM) simulations \cite{tan,saitoh}.
It is known that two shear bands are formed near the boundary and they collide to form one shear band in the center region
under a physical boundary condition.
A similar shear band formation is also observed under the Lees-Edwards boundary condition.
Such a dynamic behavior of shear bands is reproduced by a simulation of granular
hydrodynamic equations \cite{saitoh} derived from the kinetic theory for granular gases \cite{lun,dufty1,dufty2,lutsko1,lutsko2,lutsko3,jr1,jr2}.
In addition, the linear stability analyses suggest that a homogeneous state of the sheared granular flow is almost always unstable
\cite{linear1,linear2,layer1,layer2,layer3,layer4,layer5}.

Amongst many papers, it is notable that Khain found the coexistence of a solid phase and a liquid phase of granular particles
in his molecular dynamics simulation of a dense sheared granular flow \cite{khain1,khain2}.
He demonstrated the existence of a hysteresis loop of the difference of density between the boundary layer and the center region of the container
by controlling  the value of the restitution coefficient.
It should be noted that  the mechanism of an appearance of the subcritical bifurcation based on a set of hydrodynamic equations, 
differs from that observed in the jamming transition of frictional particles \cite{otsuki6}.

Recently, Shukla and Alam carried out a weakly nonlinear analysis of a plane sheared granular flow,
where they derived the Stuart-Landau equation of a disturbance amplitude under a physical boundary condition
starting from a set of granular hydrodynamic equations \cite{shukla1,shukla2,shukla3}.
They found the existence of subcritical bifurcations in both relatively dilute and dense systems, 
while  the supercritical bifurcation appears in other parameter space.
However, the Stuart-Landau equation cannot be used to explain the slow evolution of the spatial structure,
because they adopted the method by Reynolds and Potter \cite{reynolds} which does not include any spatial degrees of freedom.
We also indicate that their perturbation is based on the analysis for a finite size system,
in which the relation between the perturbation parameter and shear rate becomes unclear, because
the shear rate is fixed to unity in their paper. 

In this paper, we derive the time dependent Ginzburg-Landau (TDGL) equation 
under the Lees-Edwards boundary condition\cite{lees}  
as a  spatially dependent amplitude equation of the disturbance fields
starting from a set of granular hydrodynamic equations \cite{stuart,ss,nw,reduction1,reduction2,reduction3}.
To reduce the number of control parameters, we only focus on the behavior in the hydrodynamic limit.
We discuss the bifurcation in the hydrodynamic limit from the results of the coefficients of the TDGL equation.
The organization of this paper is as follows.
In the next section, we explain our setup and basic equations of a two dimensional sheared granular flow.
In Sec.\ref{sec:linear}, we summarize the results of the linear stability analysis.
Section \ref{sec:wna} is the main part of this paper, in which we derive the TDGL equation with the aid of the weakly nonlinear analysis.
Finally, we discuss our analysis and describe our conclusion in Sec.\ref{sec:dc}.

\section{Setup and basic equations}\label{sec:g_and_g}
\begin{figure*}
\begin{center}
\includegraphics[width=0.4\textwidth]{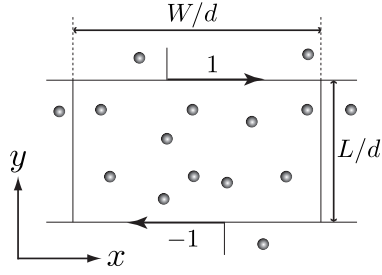}
\caption{Geometrical setup of a two dimensional
sheared granular flow under the Lees-Edwards boundary condition.
The upper and lower image cells move to the opposite direction with the dimensionless speed $1$.
The dimensionless width and height of each cell are $W/d$ and $L/d$, respectively.}
\label{fig:geometry}
\end{center}
\end{figure*}
Let us introduce our setup and basic equations.
To avoid difficulties caused by physical boundary conditions, we adopt the Lees-Edwards boundary condition, 
in which the upper and lower image cells move to the opposite direction with the speed $U/2$ \cite{lees}.
The geometry of our setup is illustrated in Fig.\ref{fig:geometry} with the Cartesian coordinate $\mathbf{x}=(x,y)$.
Because we adopt the diameter of a granular disk
$d$ and $U/2$ for the unit of length and speed, respectively, the shear rate $U/L$ is reduced to $\epsilon \equiv 2d/L$
in this dimensionless unit.
In the following, we also use the mass of a granular disk $m$ and $2d/U$ as the unit of mass and time, respectively.

We employ a set of hydrodynamic equations derived from the kinetic theory of granular gases \cite{jr1}.
Although the angular momentum and the spin temperature are included in the hydrodynamic equations,
we ignore such rotational degrees of freedom to simplify our analysis.
If the friction constant is small, this simplification can be justified, because the effect of the rotation of
granular particles during the collision can be absorbed in the normal restitution coefficient \cite{jz,yj}.

We present the derivation of the following set of dimensionless hydrodynamic equations in \ref{ap:scaling}:
\begin{eqnarray}
\left( \partial_t + \mathbf{v} \cdot \nabla \right)
\nu &=& - \nu \nabla \cdot \mathbf{v} \label{eq:hy1} \\
\nu \left( \partial_t + \mathbf{v} \cdot \nabla \right)
\mathbf{v} &=& - \nabla \cdot \mathsf{P} \label{eq:hy2} \\
\left( \nu / 2 \right) \left( \partial_t + \mathbf{v} \cdot \nabla \right)
\theta &=& - \mathsf{P} : \nabla \mathbf{v} - \nabla \cdot \mathbf{q} - \chi \label{eq:hy4}~,
\end{eqnarray}
where $\nu$, $\mathbf{v}=(u,w)$, $\theta$, $t$ and $\nabla = (\partial / \partial_x, \partial / \partial_y)$
are the area fraction, the dimensionless velocity fields, the dimensionless granular temperature,
the dimensionless time and the dimensionless gradient, respectively.
The pressure tensor $\mathsf{P}$, the heat flux  $\mathbf{q}$ and the energy dissipation rate $\chi$ are given by
\begin{eqnarray}
\mathsf{P} &=& \left[ p^\ast(\nu)\theta - \xi^\ast(\nu)\theta^{1/2}
\left( \nabla \cdot \mathbf{v} \right) \right] \delta_{ij} - \eta^\ast(\nu)\theta^{1/2} D'_{ij}~,
\label{eq:stress} \\
\mathbf{q} &=& - \kappa^\ast(\nu) \theta^{1/2} \nabla \theta - \lambda^\ast(\nu) \theta^{3/2} \nabla \nu~,
\label{eq:heat} \\
\chi &=& \frac{1-e^2}{4\sqrt{2\pi}}\nu^2 g(\nu) \theta^{1/2} \left[ 4\theta - 3\sqrt{\frac{\pi}{2}}
\theta^{1/2} \left( \nabla \cdot \mathbf{v} \right) \right]~,
\label{eq:chi}
\end{eqnarray}
respectively. Here, $D'_{ij}$ $\left(i,j = x,y \right)$ is the deviatoric part of the strain rate
\begin{equation}
D'_{ij} \equiv \frac{1}{2} \left( \nabla_j v_i + \nabla_i v_j - \delta_{ij} \nabla \cdot \mathbf{v} \right)~,
\end{equation}
and $p^\ast(\nu)\theta$, $\xi^\ast(\nu)\theta^{1/2}$, $\eta^\ast(\nu)\theta^{1/2}$, $\kappa^\ast(\nu)\theta^{1/2}$
and $\lambda^\ast(\nu)\theta^{3/2}$ are the static pressure, the bulk viscosity, the shear viscosity,
the heat conductivity and the coefficient associated with the gradient of density, respectively.
The explicit forms of them are listed in Table \ref{tab:subfunctions}, where we adopt
\begin{equation}
g(\nu) = \frac{1-7\nu/16}{\left( 1-\nu \right)^2}
\label{eq:radial}
\end{equation}
for the radial distribution function at contact which is only valid for $\nu < 0.7$ \cite{gnu4,gnu3,gnu2,gnu1}.
It should be noted that the expression of $\kappa(\nu)$ in Ref.\cite{saitoh} contains an error (see \ref{ap:scaling}).

\begin{table}
\caption{The functions in Eqs.(\ref{eq:stress})-(\ref{eq:chi}), where $e$ is the restitution coefficient.}
\label{tab:subfunctions}
\begin{tabular}{lll}
\hline\noalign{\smallskip}
$p^\ast(\nu)$ &$=$& $\frac{1}{2}\nu \left[ 1+(1+e)\nu g(\nu) \right]$ \\
$\xi^\ast(\nu)$ &$=$& $\frac{1}{\sqrt{2\pi}} (1+e)\nu^2 g(\nu)$ \\
$\eta^\ast(\nu)$ &$=$& $\sqrt{\frac{\pi}{2}} \left[ \frac{g(\nu)^{-1}}{7-3e}
+ \frac{(1+e)(3e+1)}{4(7-3e)}\nu + \left( \frac{(1+e)(3e-1)}{8(7-3e)} + \frac{1}{\pi} \right)(1+e)\nu^2 g(\nu) \right]$ \\
$\kappa^\ast(\nu)$ &$=$& $\sqrt{2\pi} \left[ \frac{g(\nu)^{-1}}{(1+e)(19-15e)}
+ \frac{3(2e^2+e+1)}{8(19-15e)}\nu + \left( \frac{9(1+e)(2e-1)}{32(19-15e)} + \frac{1}{4\pi} \right)(1+e)\nu^2 g(\nu) \right]$ \\
$\lambda^\ast(\nu)$ &$=$& $- \sqrt{\frac{\pi}{2}}\frac{3e(1-e)}{16(19-15e)}\left[ 4(\nu g(\nu))^{-1} + 3(1+e) \right] \frac{d \left( \nu^2g(\nu) \right)}{d\nu}$ \\
\noalign{\smallskip}\hline
\end{tabular}
\end{table}

\section{Linear stability analysis}\label{sec:linear}

In this section, we present the linear stability analysis of a sheared granular flow under the Lees-Edwards boundary condition.
Although the analysis is essentially same as those in the previous studies \cite{linear1,linear2,layer1,layer2,layer3,layer4,layer5},
it is necessary as the basis of the weakly nonlinear analysis.

\subsection{Linearized equation}

We introduce the hydrodynamic field and the homogeneous solution of Eqs.(\ref{eq:hy1})-(\ref{eq:hy4}) as
$\phi \equiv \left( \nu,u,w,\theta \right)^\mathrm{T}$ and $\phi_0 \equiv \left( \nu_0, \epsilon y, 0,
\theta_0 \right)^\mathrm{T}$, respectively, where the upperscript $\mathrm{T}$ represents the transposition,
$\nu_0$ is the mean area fraction and
\begin{equation}
\theta_0 = \sqrt{\frac{\pi}{2}} \frac{\epsilon^2 \eta^\ast(\nu_0)}{\left( 1-e^2 \right) \nu_0^2 g(\nu_0)}
\label{eq:theta0}
\end{equation}
is the mean granular temperature. Thus, in the hydrodynamic limit $\epsilon \ll 1$, $1-e^2$ is scaled as
$1-e^2 = \epsilon^2$ with the fixed $\theta_0$. The disturbance field is defined as
$\hat{\phi}(x,y,t) \equiv \phi - \phi_0$ which is transformed into the Fourier series
\begin{equation}
\hat{\phi}(x,y,t) = A^{\rm L}\sum_{k_{y0}} \phi^{\rm L}_{k_{y0}}e^{ik_{y0}y} 
+ A^{\rm NL} \sum_{k_x\neq 0}\sum_{k_{y0}} \phi^{\rm NL}_{\mathbf{k}(t)} e^{i \mathbf{k}(t) \cdot \mathbf{x}}~,
\label{eq:fourier}
\end{equation}
where the upperscripts L and NL respectively represent the layering mode $(k_x = 0)$ and non-layering mode $(k_x \neq 0)$, and 
$A^{\rm I}$ with ${\rm I}={\rm L}$ or NL is the amplitude. The so-called \textit{Kelvin mode} is defined as
\begin{equation}
\mathbf{k}(t) \equiv (k_x, k_y(t)) \equiv (k_x, k_{y0} - \epsilon t k_x)~,
\label{eq:Kelvin}
\end{equation}
where $k_{y0}\equiv k_y(0)$ and the coefficient $\phi^{\rm I}_{\mathbf{k}(t)}$ is defined with the imaginary unit $i$ as
\begin{equation}
\phi^{\rm I}_{\mathbf{k}(t)} = (\nu_{\mathbf{k}(t)},i u_{\mathbf{k}(t)},i w_{\mathbf{k}(t)}, \theta_{\mathbf{k}(t)})^\mathrm{T}~.
\end{equation}
We also introduce $\varphi^{\rm I}_{\mathbf{k}(t)} \equiv (\nu_{\mathbf{k}(t)},u_{\mathbf{k}(t)},w_{\mathbf{k}(t)},
\theta_{\mathbf{k}(t)})^\mathrm{T}$ for the convenience of the analysis.
If we linearize Eqs.(\ref{eq:hy1})-(\ref{eq:hy4}), $\varphi^{\rm I}_{\mathbf{k}(t)}$ satisfies
\begin{equation}
\frac{d \varphi^{\rm I}_{\mathbf{k}(t)}}{dt}=\mathcal{L}(t) \varphi^{\rm I}_{\mathbf{k}(t)}~,
\label{eq:linearized}
\end{equation}
where 
 the convective term is canceled because of the Kelvin mode Eq.(\ref{eq:Kelvin}).
The time dependent matrix $\mathcal{L}(t)$ is decomposed as
\begin{equation}
\mathcal{L}(t) = \mathcal{L}_0(k_x,k_{y0}) + t \mathcal{L}_1(k_x,k_{y0}) + t^2 \mathcal{L}_2(k_x,k_{y0})~.
\label{eq:L}
\end{equation}
The matrices $\mathcal{L}_0(k_x,k_{y0}), \mathcal{L}_1(k_x,k_{y0}), \mathcal{L}_2(k_x,k_{y0})$ are respectively given by
\begin{eqnarray}
& &\mathcal{L}_0(k_x,k_{y0}) = \nonumber \\
& &
\begin{pmatrix}
 0 &
 \nu_0 k_x &
 \nu_0 k_{y0} &
 0 \\
 \epsilon \frac{\eta'_0}{2} k_{y0} - p'_0 k_x &
 - \xi_0 k_x^2 - \frac{\eta_0}{2} k^2 &
 - \xi_0 k_x k_{y0} - \epsilon &
 \epsilon \frac{\eta_0}{4\theta_0} k_{y0} - \frac{p_0}{\theta_0} k_x \\
 \epsilon \frac{\eta'_0}{2} k_x - p'_0 k_{y0} &
 - \xi_0 k_x k_{y0} &
 - \xi_0 k_{y0}^2 - \frac{\eta_0}{2} k^2 &
 \epsilon \frac{\eta_0}{4\theta_0} k_x - \frac{p_0}{\theta_0} k_{y0} \\
 \epsilon^2 c_1 - 2 \lambda_0 k^2 &
 c_2 k_x - 2 \epsilon \eta_0 k_{y0} &
 c_2 k_{y0} - 2 \epsilon \eta_0 k_x &
 \epsilon^2 c_3 - 2 \kappa_0 k^2
\end{pmatrix}, \label{eq:L0}\\
& &
\mathcal{L}_1(k_x,k_{y0}) =
\begin{pmatrix}
 0 &
 0 &
 - \epsilon \nu_0 k_x &
 0 \\
 - \epsilon^2 \frac{\eta'_0}{2} k_x &
 \epsilon \eta_0 k_x k_{y0} &
 \epsilon \xi_0 k_x^2 &
 - \epsilon^2 \frac{\eta_0}{4\theta_0} k_x \\
 \epsilon p'_0 k_x &
 \epsilon \xi_0 k_x^2 &
 \epsilon ( 2 \xi_0 + \eta_0 ) k_x k_{y0} &
 \epsilon \frac{p_0}{\theta_0} k_x \\
 4 \epsilon \lambda_0 k_x k_{y0} &
 2 \epsilon^2 \eta_0 k_x &
 - \epsilon c_2 k_x &
 4 \epsilon \kappa_0 k_x k_{y0}
\end{pmatrix}, \label{eq:L1}\\
& &
\mathcal{L}_2(k_x,k_{y0}) =
\begin{pmatrix}
 0 &
 0 &
 0 &
 0 \\
 0 &
 - \epsilon^2 \frac{\eta_0}{2} k_x^2 &
 0 &
 0 \\
 0 &
 0 &
 - \epsilon^2 (\xi_0 + \frac{\eta_0}{2}) k_x^2 &
 0 \\
 - 2 \epsilon^2 \lambda_0 k_x^2 &
 0 &
 0 &
 - 2 \epsilon^2 \kappa_0 k_x^2
\end{pmatrix}, \label{eq:L2}
\end{eqnarray}
where $k \equiv \sqrt{k_x^2+k_{y0}^2}$ and $c_1$, $c_2$ and $c_3$ are respectively given by
\begin{eqnarray}
c_1 &=& \eta_1 - \sqrt{\frac{2}{\pi}} ( g_0 + \nu_0 g_1 ) \theta_0^{3/2}~,\\
c_2 &=& 2 p_0 - \frac{3}{4} \epsilon^2 \nu_0 g_0 \theta_0~,\\
c_3 &=& \frac{\eta_0}{2 \theta_0} - \frac{3}{\sqrt{2\pi}} \nu_0 g_0 \theta_0^{1/2}~.
\end{eqnarray}
The explicit forms of the coefficients of the Taylor expansion,
i.e. $g_0, p_0, \xi_0, \eta_0, \kappa_0, \lambda_0, g_1$ and $p'_0$
are respectively given by Eqs.(\ref{eq:Tg})-(\ref{eq:Tlambda}), (\ref{eq:Tg1234}) and (\ref{eq:p'_0}) in \ref{ap:Taylor}.

\subsection{Non-layering mode}\label{sec:nonlayer}
The solution of Eq.(\ref{eq:linearized}) is obtained by the parallel procedure in Refs.\cite{lutsko0,otsuki0,otsuki1}
for the case of \textit{the non-layering mode} $(k_x \neq 0)$. In \ref{ap:nonlayer_perturb}, we perturbatively solve
Eq.(\ref{eq:linearized}) by scaling the wave number as $\mathbf{k}(t)=\epsilon \mathbf{q}(t)$ and find the components
of $\varphi^{\rm NL}_{\mathbf{q}(t)}$ as
\begin{eqnarray}
\nu_{\mathbf{q}(t)} &=& - \frac{p_0}{\theta_0J} E^{(2)}(t)
+ \frac{\nu_0}{J} E^{(3)}(t)\cos{\omega(t)}~, \label{eq:nonlayering_decay_nu}\\
u_{\mathbf{q}(t)} &=& -\frac{\epsilon t}{\sqrt{1+(\epsilon t)^2}} E^{(1)}(t)
- \frac{1}{\sqrt{1+(\epsilon t)^2}} E^{(3)}(t) \sin\omega(t)~, \label{eq:nonlayering_decay_u}\\
w_{\mathbf{q}(t)} &=& -\frac{1}{\sqrt{1+(\epsilon t)^2}} E^{(1)}(t)
+ \frac{\epsilon t}{\sqrt{1+(\epsilon t)^2}} E^{(3)}(t) \sin\omega(t)~, \label{eq:nonlayering_decay_w}\\
\theta_{\mathbf{q}(t)} &=& \frac{p'_0}{J} E^{(2)}(t) + \frac{2p_0}{J} E^{(3)}(t)\cos{\omega(t)}~, \label{eq:nonlayering_decay_theta}
\end{eqnarray}
where we defined 
\begin{eqnarray}
E^{(1)}(t) &=& \exp\left[ - q_x^2 r_a \left( \epsilon^2 t + \frac{\epsilon^4}{3} t^3 \right) \right]~, \\
E^{(2)}(t) &=& \exp\left[ - q_x^2 r_b \left( \epsilon^2 t + \frac{\epsilon^4}{3} t^3 \right) \right]~, \\
E^{(3)}(t) &=& \exp\left[ - q_x^2 r_c \left( \epsilon^2 t + \frac{\epsilon^4}{3} t^3 \right) \right]~,
\end{eqnarray}
and the frequency
\begin{equation}
\omega(t) = \frac{q_x J}{2} \left[ \epsilon t \sqrt{1+(\epsilon t)^2}
+ \ln\left\{\epsilon t + \sqrt{1+(\epsilon t)^2}\right\} \right]~.
\end{equation}
The positive constants $J$, $r_a$, $r_b$ and $r_c$ are respectively given by Eqs.(\ref{ap:eq:J}) and
(\ref{ap:eq:rarbrc}) in \ref{ap:nonlayer_perturb}.

From Eqs.(\ref{eq:nonlayering_decay_nu})-(\ref{eq:nonlayering_decay_theta}), 
$\varphi^{\rm NL}_{\mathbf{q}(t)}$
decays to zero in the long time limit as indicated in the previous works \cite{layer1,layer3}.
Therefore, the nonlayering mode is linearly stable.
It should be, however, noted that $\varphi^{\rm NL}_{\mathbf{q}(t)}$ involving the convective effect
is only necessary for  $q_x\neq 0$ \cite{lutsko0,otsuki0,otsuki1}.
Thus, we can solve Eq.(\ref{eq:linearized}) for $q_x=0$ separately in the next subsection.

\subsection{Layering mode}\label{sec:layer}
In the case of \emph{the layering mode} ($k_x=0$), Eq.(\ref{eq:linearized}) is reduced to
the eigenvalue problem
\begin{equation}
\mathcal{L}_0(0,k_{y0}) \varphi^{\rm L}_{k_{y0}} = \sigma(k_{y0}) \varphi^{\rm L}_{k_{y0}}~, \label{eq:eigen_pb}
\end{equation}
where $\sigma(k_{y0})$ and $\varphi^{\rm L}_{k_{y0}}$ are the eigenvalue and eigenvector of $\mathcal{L}(0,k_{y0})$, respectively.
We also define the left eigenvector $\tilde{\varphi}^{\rm L}_{k_{y0}}$ as
\begin{equation}
\tilde{\varphi}^{\rm L}_{k_{y0}} \mathcal{L}_0(0,k_{y0}) = \sigma(k_{y0}) \tilde{\varphi}^{\rm L}_{k_{y0}}~. \label{eq:adjoint_pb}
\end{equation}
In \ref{ap:perturb}, we perturbatively solve Eqs.(\ref{eq:eigen_pb}) and (\ref{eq:adjoint_pb}) with the scaling 
$k_{y0}=\epsilon q$ and find the dispersion relation
\begin{equation}
\sigma(q) = \epsilon^2 ( r_2 q^2 + r_4 q^4 )~, \label{eq:dispersion}
\end{equation}
which is maximum at $q_c \equiv \sqrt{-r_2/2r_4}$, where $r_2$ and $r_4$ are given by Eqs.(\ref{ap:eq:r2}) and (\ref{ap:eq:r4}) in \ref{ap:perturb}.
The right and left eigenvectors are respectively given by
\begin{eqnarray}
\varphi^{\rm L}_q &\equiv& (\nu_q, u_q, w_q, \theta_q)^\mathrm{T}
= \left( - \frac{p_0 a_2^{(1)}}{\theta_0J}, a_1^{(1)}, \epsilon C_{\varphi_1}, \frac{p'_0 a_2^{(1)}}{J} \right)^\mathrm{T},
\label{eq:varphiq} \\
\tilde{\varphi}^{\rm L}_q &\equiv& (\tilde{\nu}_q, \tilde{u}_q, \tilde{w}_q, \tilde{\theta}_q)
= \left( \frac{p'_0\tilde{a}_1^{(1)}}{J^2 q} - \frac{2p_0\tilde{a}_2^{(1)}}{J},
\tilde{a}_1^{(1)}, \epsilon \tilde{C}_{\varphi_1}, \frac{p_0\tilde{a}_1^{(1)}}{\theta_0J^2 q}+\frac{\nu_0\tilde{a}_2^{(1)}}{J} \right),
\label{eq:adjointq}
\end{eqnarray}
where $\mathbf{a}^{(1)}=(a_1^{(1)},a_2^{(1)})^\mathrm{T}$,
$\tilde{\mathbf{a}}^{(1)}=(\tilde{a}_1^{(1)},\tilde{a}_2^{(1)})^\mathrm{T}$,
$C_{\varphi_1}$ and $\tilde{C}_{\varphi_1}$ are given by
Eqs.(\ref{eq:a12_ap}), (\ref{eq:b12_ap}), (\ref{eq:C_varphi1_ap}) and (\ref{eq:aC_varphi1_ap})
in \ref{ap:perturb}, respectively.

\begin{figure*}
\begin{center}
\includegraphics[width=0.5\textwidth]{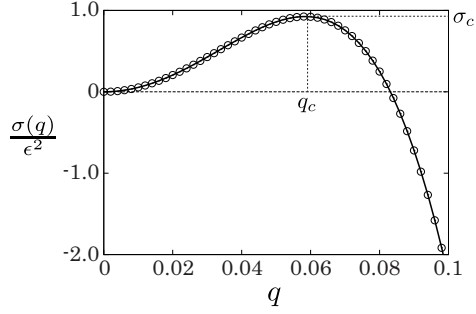}
\caption{The dispersion relation $\sigma(q)/\epsilon^2$,
where the open circles and the solid line represent the numerical results and
Eq.(\ref{eq:dispersion}), respectively. The maximum value is given by the scaled wave number $q_c=0.057$.
Here, we used $\epsilon = 0.01$, $e=0.99$ and $\nu_0=0.4$.} \label{fig:dispersion}
\end{center}
\end{figure*}

Figure \ref{fig:dispersion} shows the dispersion relation $\sigma(q)/\epsilon^2$,
where the open circles and the solid line represent the numerical results and
Eq.(\ref{eq:dispersion}), respectively. In numerical calculation, we solved the
eigenvalue problem Eq.(\ref{eq:eigen_pb}) by LAPACK \cite{lapack} with
$\epsilon = 0.01$, $e=0.99$ and $\nu_0=0.4$. In this figure, the maximum value
\begin{equation}
\sigma_c \equiv \frac{\sigma(q_c)}{\epsilon^2} = r_2 q_c^2 + r_4 q_c^4
\end{equation}
is given by $q_c=0.057$.
It should be noted that the imaginary part of $\sigma(q)$ is always zero.
\section{Weakly nonlinear analysis}\label{sec:wna}

The linear stability analysis is only useful to know whether the considered base state is stable. 
If we are interested in the structure formation after the base state becomes unstable, we need, at least, a weakly nonlinear analysis. 
Let us introduce a long time scale and long
length scales as $\tau = \epsilon^2 t$ and $\mathbf{z}=(\xi,\zeta) = \epsilon (x,y)$, respectively,
to characterize the slow and large scale evolutions of structure. 
Thus, the derivatives are replaced by
\begin{equation}
\partial_t = \epsilon^2 \partial_\tau, \hskip2em
\nabla = \epsilon (\partial_\xi, \partial_\zeta)~.
\label{eq:long_scales_d}
\end{equation}

The slow evolution of hydrodynamics variables are obtained from the evolution of the \emph{neutral solution} of the linearized equation.
The neutral solution at the most unstable mode $\mathbf{q}_c = (0,q_c)$ is given by
\begin{equation}
\hat{\phi}_\mathrm{n} = A^{\rm L}(\zeta,\tau)\phi^{\rm L}_{q_c}e^{i q_c \zeta} + \mathrm{c.c.}~,\label{eq:neutral}
\end{equation}
where each component of $\phi^{\rm L}_{q_c}$
is the corresponding one in Eq.(\ref{eq:varphiq}) at $q=q_c$,  and $\mathrm{c.c.}$ represents the complex conjugate.
It is notable that the amplitude $A^{\rm L}(\zeta,\tau)$ is independent of  $\xi$, 
because the non-layering mode $q_x\neq 0$ are linearly stable.
Thus, if we adopt the conventional approach in which the amplitude equation is obtained from the expansion around the neutral solution,
we cannot discuss the structure evolution in $\xi$ direction.

If we carry out the weakly nonlinear analysis using $\hat{\phi}_\mathrm{n}$,
the amplitude equation for $A^{\rm L}(\zeta,\tau)$ only depends on $\zeta$, but
the disturbance in the $\xi$-direction also exists  in the two-dimensional granular shear flow.
Let us try to introduce a hybrid approach to involve $\xi$ dependence in shear flow.
For this purpose, we may rewrite $\hat{\phi}(x,y,t)$ in Eq. (\ref{eq:fourier}) in the vicinity of $\mathbf{q}=\mathbf{q}_c$ as
\begin{equation}
\hat{\phi}_\mathrm{n} \simeq  a(\xi,\zeta,\tau)\phi^{\rm L}_{q_c}e^{i \mathbf{q}(\tau) \cdot \mathbf{z}} + \mathrm{c.c.}~,\label{eq:neutral1}
\end{equation}
where the wave number $\mathbf{q}(\tau)$ involve the contribution of the deviation
$\mathbf{q}_c$, i.e. $\mathbf{q}(\tau) = \mathbf{q}_c + \delta \mathbf{q}(\tau)$.
In addition, we need to include the contribution of the non-layering mode
$\phi^{\rm NL}_{\mathbf{q}(\tau)}=(\nu_{\mathbf{q}(\tau)},i u_{\mathbf{q}(\tau)},i w_{\mathbf{q}(\tau)}, \theta_{\mathbf{q}(\tau)})^\mathrm{T}$
when we are interested in the case of $q_x \neq 0$.
Thus, Eq.(\ref{eq:neutral1}) may be replaced by the \emph{hybrid solution}
\begin{eqnarray}
\hat{\phi}_\mathrm{h} &=& [ a(\xi,\zeta,\tau) \phi^{\rm L}_{q_c}
+ A^{\rm NL}(\xi,\zeta,\tau) \phi^{\rm NL}_{\mathbf{q}(\tau)} ]e^{i \mathbf{q}(\tau) \cdot \mathbf{z}} + \mathrm{c.c.} \nonumber\\
&\simeq & A(\xi,\zeta,\tau)[ \phi^{\rm L}_{q_c} + \phi^{\rm NL}_{\mathbf{q}(\tau)} ]e^{i \mathbf{q}(\tau) \cdot \mathbf{z}} + \mathrm{c.c.}
~,\label{eq:neutral2}
\end{eqnarray}
where we have used a strong assumption that the amplitudes $a(\xi,\zeta,\tau)$ and
$A^{\rm NL}(\xi,\zeta,\tau)$ are scaled by the common amplitude $A(\xi,\zeta,\tau)$.
If we carry out the weakly nonlinear analysis using $\hat{\phi}_\mathrm{h}$ instead of $\hat{\phi}_\mathrm{n}$,
the TDGL equation might depend on $\xi$.
Strictly speaking, we cannot justify the above hybrid approach between two different modes,
i.e., the layering mode and the non-layering mode.
Nevertheless, we will take into account $\xi$ dependence in the TDGL equation
phenomenologically.

Now, let us proceed the explicit calculation of weakly nonlinear analysis.
To avoid the confusion from the uncertain part in the hybrid approach,
we first derive the one-dimensional TDGL equation in Sec.\ref{sub:weakly_layering} for only the layering mode, and
give the hybrid  TDGL equation in Sec.\ref{sub:weakly_nonlayer} including the contribution from the nonlayering mode.

\subsection{Weakly nonlinear analysis of the layering mode}\label{sub:weakly_layering}

In this subsection, we derive TDGL equation as the amplitude equation for the layering mode at the most unstable wave number.
This subsection consists of three parts. In the first part, we expand the amplitude $A^{\rm L}(\zeta,\tau)$ and the matrix ${\cal L}_0(0,\epsilon q_c)$  introduced in Eq.(\ref{eq:L0}). In the second part, we will derive TDGL equation at $O(\epsilon^3)$ which is
sufficient if the bifurcation is supercritical. 
In the last part, we will present higher order calculation which is necessary if the bifurcation is subcritical.

\subsubsection{Expansions of amplitude and matrix}\label{sub:perturbation}

In this part, we prepare the expansions of the amplitude and the matrix in terms of $\epsilon$, which is necessary for the weakly nonlinear analysis.
From the straightforward calculation, $A^{\rm L}(\zeta,\tau)$ and $\mathcal{L}_0(0,\epsilon q_c)$ can be expanded as
\begin{eqnarray}
A^{\rm L}(\zeta,\tau) &=& \epsilon A^{\rm L}_1 + \epsilon^2 A^{\rm L}_2 + \epsilon^3 A^{\rm L}_3 + \dots~, \label{eq:amp_exp}\\
\mathcal{L}_0(0,\epsilon q_c) &=& \epsilon \mathcal{M}_{1}
+ \epsilon^2 \mathcal{M}_{2} + \dots~, \label{eq:LM12}
\end{eqnarray}
where the matrix $\mathcal{L}_0$ is introduced in Eq.(\ref{eq:L0}) and
\begin{equation}
\mathcal{M}_1 =
\begin{pmatrix}
 0 & 0 & \nu_0 q_c & 0 \\
 0 & 0 & - 1 & 0 \\
 - p'_0 q_c & 0 & 0 & - \frac{p_0}{\theta_0} q_c \\
 0 & 0 & 2 p_0 q_c & 0
\end{pmatrix}~,
\end{equation}
\begin{equation}
\mathcal{M}_2 =
\begin{pmatrix}
 0 & 0 & 0 & 0 \\
 \frac{\eta'_0}{2} q_c & - \frac{\eta_0}{2} q_c^2 & 0 & \frac{\eta_0}{4\theta_0} q_c \\
 0 & 0 & - ( \xi_0 + \frac{\eta_0}{2} ) q_c^2 & 0 \\
 c_1 & - 2 \eta_0 q_c & 0 & c_3 - 2 \kappa_0 q_c^2
\end{pmatrix}~.
\end{equation}
Substituting Eqs.(\ref{eq:neutral}) and (\ref{eq:amp_exp}) into Eqs.(\ref{eq:hy1})-(\ref{eq:hy4})
and collecting the order of $\epsilon$, we can obtain a series of terms of equations.

Multiplying the left zero-eigenvector of ${\cal L}_0(0,\epsilon q_c)$, we will obtain the amplitude equation.
It is not easy to obtain the left zero-eigenvector in general, but fortunately $\tilde{\varphi}^{\rm L}_{q_c}$
introduced in Eq.(\ref{eq:adjointq}) plays a role of the zero-eigenvector in the limit $\epsilon \to 0$
thanks to Eqs.(\ref{eq:eigen_pb}) and (\ref{eq:adjoint_pb}).

\subsubsection{The TDGL equation at $O(\epsilon^3)$}
The first nonzero terms appear at $O(\epsilon^2)$, where the coefficient of $e^{i q_c \zeta}$ satisfies
\begin{equation}
\mathcal{M}_{1} \varphi_{q_c}^{\rm L}=0 , \label{eq:O2}
\end{equation}
where $\varphi_{q_c}^{\rm L}$ is introduced in Eq.(\ref{eq:varphiq}).
At $O(\epsilon^3)$, the coefficient of $e^{i q_c \zeta}$ satisfies
\begin{equation}
\varphi^{\rm L}_{q_c} \partial_\tau A^{\rm L}_1 = \mathcal{M}_2 \varphi^{\rm L}_{q_c} A^{\rm L}_1
+ \mathcal{D} \partial_\zeta^2 A^{\rm L}_1 + \mathcal{N}_3 A^{\rm L}_1 |A^{\rm L}_1|^2~, \label{eq:O3}
\end{equation}
where $\mathcal{D}$ and $\mathcal{N}_3$ are given by
\begin{equation}
\mathcal{D} =
\begin{pmatrix}
 0 \\
 \eta_0 u_{q_c}/2 \\
 (\xi_0 + \eta_0/2)w_{q_c} \\
 2 \kappa_0 \theta_{q_c}
\end{pmatrix}, \hskip1.5em
\mathcal{N}_3 =
\begin{pmatrix}
 0  \\
 0 \\
 - p'_2 \nu_{q_c}^3 - (p'_1+p_2) \nu_{q_c}^2 \theta_{q_c}/\theta_0 \\
 2 \nu_{q_c} w_{q_c} ( p_1 \theta_{q_c} / \theta_0 + p_2 \nu_{q_c} )
\end{pmatrix}~, \label{eq:N3}
\end{equation}
respectively. 

If we multiply the left zero-eigenvector $\tilde{\varphi}^{\rm L}_{q_c}$ to Eq.(\ref{eq:O3})
introduced in Eq.(\ref{eq:adjointq}), we obtain the TDGL equation:
\begin{equation}
\partial_\tau A^{\rm L}_1 = \sigma_c A^{\rm L}_1 + d \partial_\zeta^2 A^{\rm L}_1 + \beta A^{\rm L}_1 |A^{\rm L}_1|^2~,
\label{eq:GL30}
\end{equation}
where we have used the normalized condition $\tilde{\varphi}^{\rm L}_{q_c} \varphi^{\rm L}_{q_c} = 1$, and $d$ and $\beta$ are given by
\begin{eqnarray}
d &=& \frac{\eta_0}{2} (\tilde{u}_{q_c} u_{q_c} + \tilde{w}_{q_c} w_{q_c})
+ \xi_0 \tilde{w}_{q_c} w_{q_c} + 2 \kappa_0 \tilde{\theta}_{q_c} \theta_{q_c}~, \label{eq:all_d} \\
\beta &=& 2 \tilde{\theta}_{q_c} \nu_{q_c} w_{q_c}
\left( p_2 \nu_{q_c} + \frac{p_1}{\theta_0} \theta_{q_c} \right)
- \tilde{w}_{q_c} \left( p'_2 \nu_{q_c}^3 + \frac{p'_1+p_2}{\theta_0} \nu_{q_c}^2 \theta_{q_c} \right)~,
\label{eq:all_beta}
\end{eqnarray} 
respectively.

Substituting Eqs.(\ref{eq:varphiq}) and (\ref{eq:adjointq}) to
Eqs.(\ref{eq:all_d}) and (\ref{eq:all_beta}), the leading terms of $\epsilon$ give
\begin{equation}
d = \bar{d}~,\hskip2em \beta = \epsilon \bar{\beta}~,
\label{eq:leading}
\end{equation}
where $\bar{d}$ and $\bar{\beta}$ are listed in Table \ref{tab:TDGL_constants}. It is notable that the coefficient
$\beta$ becomes higher order of $\epsilon$. 
Therefore,
we need to rescale the amplitude as
\begin{equation}
\overline{A^{\rm L}}_1(\zeta,\tau) = \epsilon^{1/2} A^{\rm L}_1(\zeta,\tau) \label{eq:A_bar}
\end{equation}
and the TDGL equation for $\overline{A^{\rm L}}_1(\zeta,\tau)$ is reduced to
\begin{equation}
\partial_{\tau} \overline{A^{\rm L}}_1 = \sigma_c \overline{A^{\rm L}}_1 + \bar{d} \partial_{\zeta}^2 \overline{A^{\rm L}}_1
+ \bar{\beta} \overline{A^{\rm L}}_1 |\overline{A^{\rm L}}_1|^2~. \label{eq:red_TDGL3_layer}
\end{equation}
The scaling relation Eq.(\ref{eq:A_bar}) indicates that the amplitude of $\hat{\phi}$ is extended as
\begin{equation}
\epsilon^{1/2} \overline{A^{\rm L}}_1(\zeta,\tau) + \epsilon^{3/2} \overline{A^{\rm L}}_2(\zeta,\tau)
+ \epsilon^{5/2} \overline{A^{\rm L}}_3(\zeta,\tau) + \dots~, \label{eq:A_bar_expansion}
\end{equation}
where $\overline{A^{\rm L}}_j = \epsilon^{1/2} A^{\rm L}_j$ ($j=2,3,\dots$).
Thus, Eq.(\ref{eq:A_bar_expansion}) converges to zero in the limit $\epsilon \rightarrow 0$.

Let us compare Eq.(\ref{eq:leading}) with the numerical result, where we solve Eq.(\ref{eq:eigen_pb})
by LAPACK and calculate $\beta$ from Eq.(\ref{eq:all_beta}). We find
$\sigma_c$ and $\bar{d}$ are always positive and Eq.(\ref{eq:leading}) perfectly agrees with the numerical results
(Fig.\ref{fig:beta_gamma}). We find $\bar{\beta}<0$ in $0<\nu_0<0.245$, thus,
a supercritical bifurcation can be observed in the dilute regime. On the other hand, 
and a subcritical bifurcation , i.e. $\bar{\beta}>0$ appears in $\nu_0>0.245$.

\begin{figure*}
\begin{center}
\includegraphics[width=0.4\textwidth]{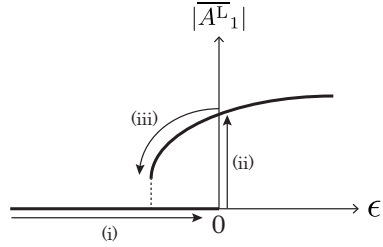}
\caption{A schematic image of the subcritical bifurcation of $|\overline{A^{\mathrm L}}_1|$,
where a hysteresis loop is realized by the paths (i), (ii) and (iii).}
\label{fig:hysteresis}
\end{center}
\end{figure*}

It should be noted that there is no hysteresis behavior even for the subcritical bifurcation.
Figure \ref{fig:hysteresis} is a schematic image of the subcritical bifurcation of $|\overline{A^{\rm L}}_1|$,
where a hysteresis loop is realized by the paths (i), (ii) and (iii).
Because we restrict our interest to the case of $\epsilon>0$ from the definition, the paths (i) and (iii) cannot exist.
Therefore, such a hysteresis behavior cannot be observed in the hydrodynamic limit.
\subsubsection{Higher order expansions}
Because of $\bar{\beta}>0$ in $\nu_0>0.245$, we need to proceed our calculation to the higher order expansions.
At $O(\epsilon^4)$ and $O(\epsilon^5)$, the coefficients of $e^{i q_c \zeta}$ satisfy
\begin{eqnarray}
& & \varphi^{\rm L}_{q_c} \partial_\tau A^{\rm L}_2 = \mathcal{M}_2 \varphi^{\rm L}_{q_c} A^{\rm L}_2
+ \mathcal{D} \partial_\zeta^2 A^{\rm L}_2 + \mathcal{N}_3 ( {A^{\rm L}_1}^2 {A^{\rm L}_2}^\ast + 2 A^{\rm L}_2 |A^{\rm L}_1|^2 )~, \label{eq:O4}\\
& & \varphi^{\rm L}_{q_c} \partial_\tau A^{\rm L}_3 = \mathcal{M}_2 \varphi^{\rm L}_{q_c} A^{\rm L}_3
+ \mathcal{D} \partial_\zeta^2 A^{\rm L}_3 + \mathcal{N}_3 ({A^{\rm L}_1}^\ast {A^{\rm L}_2}^2 + 2 A^{\rm L}_1 |A^{\rm L}_2|^2 + 
{A^{\rm L}_1}^2 {A^{\rm L}_3}^\ast
+ 2|A^{\rm L}_1|^2 A^{\rm L}_3 ) \nonumber \\
& & + \mathcal{N}_5 A^{\rm L}_1 |A^{\rm L}_1|^4
+ \mathcal{B} ({A^{\rm L}_1}^2 \partial_\zeta^2 {A^{\rm L}_1}^\ast + 2 |A^{\rm L}_1|^2 \partial_\zeta^2 A^{\rm L}_1 )
+ \mathcal{C} \{{A^{\rm L}_1}^\ast (\partial_\zeta A^{\rm L}_1)^2 + 2 A^{\rm L}_1 |\partial_\zeta A^{\rm L}_1|^2 \}, \label{eq:O5}
\end{eqnarray}
respectively, where ${A^{\rm L}_j}^\ast$~$(j=1,2,3)$ represents the complex conjugate of $A^{\rm L}_j$ and
\begin{equation}
\mathcal{N}_5 =
\begin{pmatrix}
 0  \\
 0 \\
 - 2 p'_3 \nu_{q_c}^4 \theta_{q_c}/\theta_0 \\
 4 p_3 \nu_{q_c}^3 \theta_{q_c} w_{q_c}/\theta_0
\end{pmatrix}~. \label{eq:N5}
\end{equation}
Although the vectors $\mathcal{B}$ and $\mathcal{C}$ can be written explicitly,
we do not need these analytic forms in later discussion.

Let us introduce the \emph{envelope function}
\begin{equation}
\tilde{A^{\rm L}}(\zeta,\tau) \equiv A^{\rm L}_1(\zeta,\tau) + \epsilon A^{\rm L}_2(\zeta,\tau) + \epsilon^2 A^{\rm L}_3(\zeta,\tau)~,
\label{eq:gen_amp}
\end{equation}
which is used by many authors to derive higher order amplitude equations \cite{genamp1,genamp2,genamp3}.
Summing up Eqs.(\ref{eq:O3}), (\ref{eq:O4}) and (\ref{eq:O5}), we obtain
\begin{eqnarray}
& &
\varphi^{\rm L}_{q_c} \partial_\tau \tilde{A^{\rm L}} = \mathcal{M}_2 \varphi^{\rm L}_{q_c} \tilde{A^{\rm L}}
+ \mathcal{D} \partial_\zeta^2 \tilde{A^{\rm L}} + \mathcal{N}_3 \tilde{A^{\rm L}} |\tilde{A^{\rm L}}|^2
+ \epsilon^2 \mathcal{N}_5 \tilde{A^{\rm L}} |\tilde{A^{\rm L}}|^4 \nonumber \\
& & + \epsilon^2 \Big[ \mathcal{B} (\tilde{A^{\rm L}}^2 \partial_\zeta^2 \tilde{A^{\rm L}}^\ast + 2 |\tilde{A^{\rm L}}|^2 \partial_\zeta^2 \tilde{A^{\rm L}} )
+ \mathcal{C} \{\tilde{A^{\rm L}}^\ast (\partial_\zeta \tilde{A^{\rm L}})^2 + 2 \tilde{A^{\rm L}} |\partial_\zeta \tilde{A^{\rm L}}|^2 \} \Big]~.
\label{eq:O3+O4+O5}
\end{eqnarray}
Then, multiplying $\tilde{\varphi}^{\rm L}_{q_c}$ to Eq.(\ref{eq:O3+O4+O5}) we find
\begin{eqnarray}
& &
\partial_\tau \tilde{A^{\rm L}} = \sigma_c \tilde{A^{\rm L}} + d \partial_\zeta^2 \tilde{A^{\rm L}}
+ \beta \tilde{A^{\rm L}} |\tilde{A^{\rm L}}|^2 + \epsilon^2 \gamma \tilde{A^{\rm L}} |\tilde{A^{\rm L}}|^4 \nonumber \\
& & + \epsilon^2 \Big[ b (\tilde{A^{\rm L}}^2 \partial_\zeta^2 \tilde{A^{\rm L}}^\ast + 2 |\tilde{A^{\rm L}}|^2 \partial_\zeta^2 \tilde{A^{\rm L}} )
+ c \{\tilde{A^{\rm L}}^\ast (\partial_\zeta \tilde{A^{\rm L}})^2 + 2 \tilde{A^{\rm L}} |\partial_\zeta \tilde{A^{\rm L}}|^2 \} \Big],
\label{eq:GL5}
\end{eqnarray}
where $b \equiv \tilde{\varphi}^{\rm L}_{q_c} \mathcal{B}$, $c \equiv \tilde{\varphi}^{\rm L}_{q_c} \mathcal{C}$ and
\begin{eqnarray}
\gamma &=& \frac{2}{\theta_0} \nu_{q_c}^3 \theta_{q_c}
( 2p_3 \tilde{\theta}_{q_c} w_{q_c} - p'_3 \tilde{w}_{q_c} \nu_{q_c} )~. \label{eq:all_gamma}
\end{eqnarray}
Substituting Eqs.(\ref{eq:varphiq}) and (\ref{eq:adjointq}) to Eq.(\ref{eq:all_gamma}),
the leading terms of $\epsilon$ give
\begin{equation}
\gamma = \epsilon \bar{\gamma}~, \hskip2em
b = \epsilon^{2} \bar{b}~, \hskip2em
c = \epsilon^{2} \bar{c}~,
\label{eq:leading5}
\end{equation}
where $\bar{\gamma}$ is given in Table \ref{tab:TDGL_constants}. Although $\bar{b}$
and $\bar{c}$ can be written explicitly, we do not need such analytic forms in later discussion.
It is notable that the coefficient $\gamma$ becomes higher order of $\epsilon$ by substituting Eqs.(\ref{eq:varphiq}) and (\ref{eq:adjointq}).
Thus,  the rescaled envelope function $\check{A^{\rm L}}(\zeta,\tau) = \epsilon^{1/2} \tilde{A^{\rm L}}(\zeta,\tau)$ satisfies
\begin{eqnarray}
& &
\partial_\tau \check{A^{\rm L}} =
\sigma_c \check{A^{\rm L}} + \bar{d} \partial_\zeta^2 \check{A^{\rm L}} + \bar{\beta} \check{A^{\rm L}} |\check{A^{\rm L}}|^2 + \epsilon \bar{\gamma} \check{A^{\rm L}} |\check{A^{\rm L}}|^4 \nonumber \\
& & + \epsilon^3 \Big[ \bar{b} (\check{A^{\rm L}}^2 \partial_\zeta^2 \check{A^{\rm L}}^\ast + 2 |\check{A^{\rm L}}|^2 \partial_\zeta^2 \check{A^{\rm L}} )
+ \bar{c} \{\check{A^{\rm L}}^\ast (\partial_\zeta \check{A^{\rm L}})^2 + 2 \check{A^{\rm L}} \partial_\zeta \check{A^{\rm L}} \partial_\zeta \check{A^{\rm L}}^\ast \} \Big]~,
\label{eq:GL5_red}
\end{eqnarray}
where the TDGL equation including the term of $\check{A^{\rm L}} |\check{A^{\rm L}}|^4$ is given in the first line.

\begin{figure*}
\begin{center}
\includegraphics[width=\textwidth]{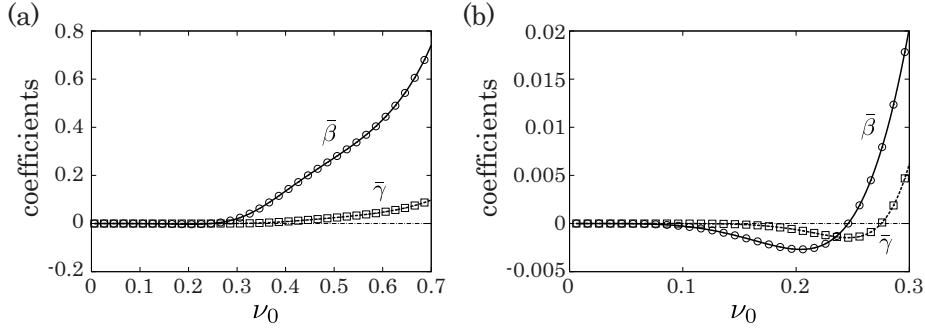}
\caption{The TDGL coefficients $\bar{\beta}$ and $\bar{\gamma}$ for (a) $0<\nu_0<0.7$
and (b) the dilute regime $0<\nu_0<0.3$, where the open circles and the open squares
represent the numerical results of $\bar{\beta}$ and $\bar{\gamma}$, respectively.
The solid and the broken lines represent the analytic results of $\bar{\beta}$ and
$\bar{\gamma}$, respectively. Here, we used $\epsilon = 0.01$.}
\label{fig:beta_gamma}
\end{center}
\end{figure*}

Let us compare Eq.(\ref{eq:leading5}) with the numerical result, where we solve Eq.(\ref{eq:eigen_pb})
by LAPACK and calculate $\gamma$ from Eq.(\ref{eq:all_gamma}).
Figure \ref{fig:beta_gamma} exhibits a complete agreement between Eqs.(\ref{eq:all_gamma}) and (\ref{eq:leading5}).
From this result, for $0.245<\nu_0<0.275$, the growth of disturbance is inhibited by the nonlinear term
$\epsilon \bar{\gamma} \check{A^{\rm L}} |\check{A^{\rm L}}|^4$ and finite steady amplitude can be observed.
For $\nu_0>0.275$, we need to calculate higher order expansions, however, it is too complicated to perform in this paper.
\subsection{Hybrid approach of weakly nonlinear analysis}\label{sub:weakly_nonlayer}

In the previous subsection, we have obtained the amplitude equation for the layering mode. 
The derivation is straightforward and the obtained amplitude equation has a reasonable form.
The equation, however, only depends on $\zeta$, and thus, we cannot discuss the spatial structure along the mean flow direction $\xi$.
To improve this unsatisfied situation, we adopt the hybrid approach as mentioned,
though it is hard to justify this approach.
Fortunately the contribution of $\varphi_{\mathbf{q}(\tau)}^{\rm NL}$ except for the diffusive mode
becomes irrelevant as time goes on. Thus, $\tilde{\varphi}^{\rm L}_{q_c}$ still can play a role of the left zero-eigenvector in our calculation. 

Let us expand the amplitude of $\hat{\phi}_\mathrm{h}$ into the series of $\epsilon$ as
\begin{equation}
A(\xi,\zeta,\tau) = \epsilon A_1(\xi,\zeta,\tau) + \epsilon^2 A_2(\xi,\zeta,\tau) + \epsilon^3 A_3(\xi,\zeta,\tau) + \dots~.
\end{equation}
If we use $\hat{\phi}_\mathrm{h}$ instead of $\hat{\phi}_\mathrm{n}$ and multiplying 
the approximate left zero-eigenvector $\tilde{\varphi}^{\rm L}_{q_c}$, we obtain the TDGL equation of $A_1(\xi,\zeta,\tau)$ as
\begin{equation}
\partial_\tau A_1 = \sigma_c A_1 + d_1(\tau) \partial_\xi^2 A_1 + d_2(\tau) \partial_\xi \partial_\zeta A_1
+ d \partial_\zeta^2 A_1 + \beta A_1 |A_1|^2~, \label{eq:GL3}
\end{equation}
where we introduced the time dependency in the diffusion constants
\begin{eqnarray}
d_1(\tau) &=& \{\frac{\eta_0}{2} (\tilde{u}_{q_c} u_{\mathbf{q}(\tau)} + \tilde{w}_{q_c} w_{\mathbf{q}(\tau)})
+ \xi_0 \tilde{u}_{q_c} u_{\mathbf{q}(\tau)} + 2 \kappa_0 \tilde{\theta}_{q_c} \theta_{\mathbf{q}(\tau)} \} \nonumber\\
& &/(1+\tilde{\varphi}^{\rm L}_{q_c} \varphi^{\rm NL}_{\mathbf{q}(\tau)}), \label{eq:all_d1} \\
d_2(\tau) &=& \xi_0 (\tilde{u}_{q_c} w_{\mathbf{q}(\tau)} + \tilde{w}_{q_c} u_{\mathbf{q}(\tau)})
/(1+\tilde{\varphi}^{\rm L}_{q_c} \varphi^{\rm NL}_{\mathbf{q}(\tau)}), \label{eq:all_d2}
\end{eqnarray}
which decay to zero because of Eqs.(\ref{eq:nonlayering_decay_nu})-(\ref{eq:nonlayering_decay_theta}).
To obtain (\ref{eq:GL3}) we ignore contributions from $\varphi_{\mathbf{q}(\tau)}^{\rm NL}$ except for the diffusion coefficient,
because they exponentially decay to zero.

Substituting Eqs.(\ref{eq:nonlayering_decay_nu})-(\ref{eq:nonlayering_decay_theta}), (\ref{eq:varphiq}) and (\ref{eq:adjointq}),
and the introduction of the scaled amplitude
\begin{equation}
\bar{A}_1(\xi,\zeta,\tau) \equiv \epsilon^{1/2} A_1(\xi,\zeta,\tau)~,
\end{equation}
we find the TDGL equation of $\bar{A}_1(\xi,\zeta,\tau)$:
\begin{equation}
\partial_\tau \bar{A}_1 = \sigma_c \bar{A}_1 + \bar{d}_1(\tau) \partial_\xi^2 \bar{A}_1 + \bar{d}_2(\tau) \partial_\xi \partial_\zeta \bar{A}_1
+ \bar{d} \partial_\zeta^2 \bar{A}_1 + \bar{\beta} \bar{A}_1 |\bar{A}_1|^2~, \label{eq:hybrid_GL3_red}
\end{equation}
where the explicit forms of $\bar{d}_1(\tau)$ and $\bar{d}_2(\tau)$ are listed in Table \ref{tab:TDGL_constants}.

From the parallel argument to obtaining Eq.(\ref{eq:GL5_red}),  the scaled envelope function introduced
\begin{equation}
\check{A}(\xi,\zeta,\tau) \equiv \epsilon^{1/2} \{A_1(\xi,\zeta,\tau)+\epsilon A_2(\xi,\zeta,\tau)+\epsilon^2 A_3(\xi,\zeta,\tau)\}~,
\end{equation}
satisfies  the TDGL equation of $\check{A}(\xi,\zeta,\tau)$:
\begin{equation}
\partial_\tau \check{A} =
\sigma_c \check{A} + d_1(\tau) \partial_\xi^2 \check{A} + d_2(\tau) \partial_\xi \partial_\zeta \check{A}
+ \bar{d} \partial_\zeta^2 \check{A} + \bar{\beta} \check{A} |\check{A}|^2 + \epsilon \bar{\gamma} \check{A} |\check{A}|^4~, \label{eq:hybrid_GL5_red}
\end{equation}
where we truncated the higher order terms of $O(\epsilon^3)$ in Eq.(\ref{eq:hybrid_GL5_red}).

\begin{figure*}
\begin{center}
\includegraphics[width=0.5\textwidth]{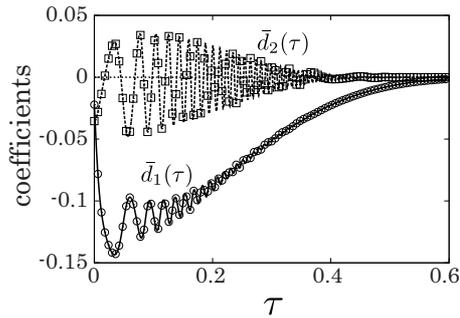}
\caption{The time development of $\bar{d}_1(\tau)$ and $\bar{d}_2(\tau)$.
The open circles and open squares represent the numerical results of $\bar{d}_1(\tau)$ and $\bar{d}_2(\tau)$, respectively.
The solid and broken lines respectively represent the analytic results of $\bar{d}_1(\tau)$ and $\bar{d}_2(\tau)$.
Here, we used $\epsilon = 0.01$, $e=0.99$ and $\nu_0=0.4$.}
\label{fig:diffusion}
\end{center}
\end{figure*}

Figure \ref{fig:diffusion} shows the time evolution of $\bar{d}_1(\tau)$ and $\bar{d}_2(\tau)$,
where the analytic results perfectly agree with the numerical calculation of Eqs.(\ref{eq:all_d1}) and (\ref{eq:all_d2}) based on LAPACK.
Because $\bar{d}_1(\tau)$ and $\bar{d}_2(\tau)$ decay to zero, Eqs.(\ref{eq:hybrid_GL3_red}) and
(\ref{eq:hybrid_GL5_red}) respectively reduce to Eq.(\ref{eq:red_TDGL3_layer}) and the first line of Eq.(\ref{eq:GL5_red}) in the long time limit.
This result is consistent with the observation in the simulation\cite{saitoh} in which
the shear band finally becomes parallel to mean-flow direction, though the mathematical justification of our hybrid approach is difficult.
%
\begin{table}
\caption{The explicit forms of $\bar{d}$, $\bar{\beta}$, $\bar{\gamma}$, $\bar{d}_1(\tau)$ and $\bar{d}_2(\tau)$.}
\label{tab:TDGL_constants}
\begin{tabular}{lll}
\hline\noalign{\smallskip}
$\bar{d}$ &$=$& $\frac{\eta_0}{2} \tilde{a}_1^{(1)} a_1^{(1)}
+ \frac{2p'_0\kappa_0}{\theta_0 J^3 q_c}\left(p_0\tilde{a}_1^{(1)} + \nu_0\theta_0Jq_c\tilde{a}_2^{(1)}\right)a_2^{(1)}$ \\
$\bar{\beta}$ &$=$& $\frac{p_0}{\theta_0^3J^4 q_c} \Big[ 2C_{\varphi_1}(p_0p_2-p'_0p_1)
\left( p_0\tilde{a}_1^{(1)}+\nu_0\theta_0Jq_c\tilde{a}_2^{(1)} \right)
+Jq_c\tilde{C}_{\varphi_1}p_0 \{ p_0p'_2-(p'_1+p_2)p'_0 \}a_2^{(1)} \Big]{a_2^{(1)}}^2$ \\
$\bar{\gamma}$ &$=$& $- \frac{2p_0^3p'_0}{\theta_0^5J^6q_c}
\left\{ 2p_3C_{\varphi_1} \left( p_0\tilde{a}_1^{(1)}+\nu_0\theta_0Jq_c\tilde{a}_2^{(1)} \right)
+Jq_cp_0p'_3\tilde{C}_{\varphi_1}a_2^{(1)} \right\}{a_2^{(1)}}^4$ \\
$\bar{d}_1(\tau)$ &$=$& $\left\{ \left( \frac{\eta_0}{2}+\xi_0 \right)\tilde{a}_1^{(1)} u_{\mathbf{q}(\tau)}
+ \frac{2\kappa_0}{\theta_0J^2q_c}(p_0 \tilde{a}_1^{(1)}+\nu_0\theta_0Jq_c\tilde{a}_2^{(1)}) \theta_{\mathbf{q}(\tau)} \right\}$\\
 & & $/\left\{ 1+\frac{1}{J^2q_c}(p'_0\tilde{a}_1^{(1)}-2p_0Jq_c\tilde{a}_2^{(1)})
\nu_{\mathbf{q}(\tau)} + \tilde{a}_1^{(1)}u_{\mathbf{q}(\tau)}
+ \frac{1}{\theta_0J^2q_c}(p_0\tilde{a}_1^{(1)} + \nu_0\theta_0Jq_c\tilde{a}_2^{(1)})\theta_{\mathbf{q}(\tau)} \right\}$ \\
$\bar{d}_2(\tau)$ &$=$& $\xi_0\tilde{a}_1^{(1)} w_{\mathbf{q}(\tau)}
/\left\{ 1+\frac{1}{J^2q_c}(p'_0\tilde{a}_1^{(1)}-2p_0Jq_c\tilde{a}_2^{(1)})
\nu_{\mathbf{q}(\tau)} + \tilde{a}_1^{(1)}u_{\mathbf{q}(\tau)}
+ \frac{1}{\theta_0J^2q_c}(p_0\tilde{a}_1^{(1)} + \nu_0\theta_0Jq_c\tilde{a}_2^{(1)})\theta_{\mathbf{q}(\tau)} \right\}$ \\
\noalign{\smallskip}\hline
\end{tabular}
\end{table}
\section{Discussion and Conclusion}\label{sec:dc}
Let us compare our results with the previous studies \cite{shukla1,shukla2,shukla3}.
The previous studies only derived Stuart-Landau equation which is independent of the position, 
while we obtain TDGL equation which can discuss the slow evolution of long-wave spatial structure.
We have demonstrated that the coefficient of $\check{A} |\check{A}|^4$ can be calculated
explicitly based on a systematic perturbation method in terms of small $\epsilon$, 
which has not been achieved by previous studies.
The appearance condition of the subcritical bifurcation is slightly different from that of the previous studies.
We believe, however, that  the result becomes similar
to that of the previous studies,
 if we analyze a finite size system around most unstable mode.
On the other hand, it is hard to justify our hybrid approach to
introduce the time dependent diffusion coefficients $d_1(\tau)$ and $d_2(\tau)$ in
the TDGL equations Eqs.(\ref{eq:hybrid_GL3_red}) and (\ref{eq:hybrid_GL5_red}), though the result seems to be reasonable.
The mathematical justification of the hybrid approach will be our future work.

In conclusion, we have derived the TDGL equation starting from a set of granular hydrodynamic equations.
From our results, we find the homogeneous state is always unstable and a supercritical bifurcation
can be observed in the dilute regime $0<\nu_0<0.245$. On the other hand, a subcritical bifurcation is
predicted in $\nu_0>0.245$ and we find the amplitude of disturbance can be converged by
the nonlinear term $\epsilon \bar{\gamma} \check{A} |\check{A}|^4$ in the range $0.245<\nu_0<0.275$.
In the case of $\nu_0>0.275$, higher order expansions are necessary, however, such calculations should
be performed in a future work.
%
%
\begin{acknowledgements}
We would like to thank M. Otsuki, H. Nakao and M. Alam for fruitful discussions,
and S. Luding for his encouragement of the initiation of this study.
This work was supported by the Global COE Program "The Next Generation of Physics, Spun from Universality \& Emergence"
from the Ministry of Education, Culture, Sports, Science and Technology (MEXT) of Japan,
the Research Fellowship of the Japan Society for the Promotion of Science for Young Scientists (JSPS),
and the Grant-in-Aid of MEXT (Grants No.21015016, 21540384 and 21.1958).
\end{acknowledgements}

\appendix
\def\thesection{Appendix \Alph{section}}

\section{Derivation of the coefficients in Table \ref{tab:subfunctions}}
\label{ap:scaling}

In this Appendix, we derive the coefficients in Table \ref{tab:subfunctions}
by using the dimensionless quantities based on the kinetic theory \cite{jr1}.
At first, the energy sources $\chi_{\alpha\alpha}$ and $\chi_3$ for smooth disks
($\beta=-1$ in Ref.\cite{jr1}) are
\begin{equation}
\chi_{\alpha\alpha} = -\frac{\xi (1-e)}{2d^2}
\left[ 8T-3\pi^{1/2} d T^{1/2} (\mathbf{\nabla}\cdot\mathbf{u}) \right] \label{eq:chi1}
\end{equation}
and $\chi_{33} = 0$, where $d$, $T\equiv m \langle (\mathbf{c}-\mathbf{u})^2\rangle/2$
and $\mathbf{u}=U\mathbf{v}/2$ are the diameter of a disk, the granular temperature and the velocity field, respectively.
It should be noted that we adopt the different definition for the granular temperature $T$ from Ref.\cite{jr1} to keep the dimension of the energy.
In Eq.(\ref{eq:chi1}), the bulk viscosity $\xi$ is given by 
\begin{equation}
\xi = \frac{4m}{\pi^{3/2} d} (1+e) \nu^2 g(\nu) T^{1/2}~. \label{eq:alpha}
\end{equation}
where $\nu$ and $g(\nu)$ is the area fraction and the radial distribution function at contact, respectively.
Thus, the factor of Eq.(\ref{eq:chi1}) is given by
\begin{equation}
-\frac{\xi (1-e)}{2\sigma^2} = -\frac{2m \nu^2 (1-e^2)}{\pi^{3/2} d^3} g(\nu) T^{1/2}~.
\end{equation}
If we introduce the mass density of the system $\rho=4m \nu /(\pi d^2)$ and the mass density of a disk
$\rho_p = 4m/(\pi d^2)$, Eq.(\ref{eq:chi1}) is reduced to
\begin{equation}
\chi_{\alpha\alpha} = -\frac{1-e^2}{2 \rho_p \pi^{1/2} d} \rho^2 g(\nu) T^{1/2}
\left[ 8T-3\pi^{1/2} d T^{1/2} (\mathbf{\nabla}\cdot\mathbf{u}) \right]~,
\end{equation}
and the energy loss rate is given by
\begin{equation}
\chi = - \frac{\chi_{\alpha\alpha}}{2} = \frac{1-e^2}{4 \rho_p \pi^{1/2} d} \rho^2 g(\nu) T^{1/2}
\left[ 8T-3\pi^{1/2} d T^{1/2} (\mathbf{\nabla}\cdot\mathbf{u}) \right]~.
\end{equation}
The pressure tensor is given by $P_{ij} = \rho T \delta_{ij} + \rho a_{ij} + \Theta_{ij}$, where
\begin{eqnarray}
\rho a_{ij} &=& - \frac{2mT^{1/2}}{\pi^{1/2} d g(\nu) (5-3r) }
\left[ 1 + \nu g(\nu) (3r-2)r \right] D'_{ij}~, \\
\Theta_{ij} &=& (2\rho T \nu g(\nu) r - \xi \mathbf{\nabla}\cdot\mathbf{u}) \delta_{ij}
- \xi D'_{ij} + \nu g(\nu) r \rho a_{ij}~,
\end{eqnarray}
with $r=(1+e)/2$. Then, we find
\begin{equation}
P_{ij} = \left[ p - \xi \mathbf{\nabla}\cdot\mathbf{u} \right] \delta_{ij}
- \xi D'_{ij} + \left[ 1 + r \nu g(\nu) \right] \rho a_{ij}~, \label{eq:ptensor1}
\end{equation}
where the static pressure is given by $p = \rho T \left[ 1 + (1+e) \nu g(\nu) \right]$.
The second and third terms on the right-hand-side of Eq.(\ref{eq:ptensor1}) can be rewritten as
$\xi D'_{ij} - \left[ 1 + r \nu g(\nu) \right] \rho a_{ij} \equiv \eta D'_{ij}$, where the shear viscosity $\eta$ is given by
\begin{equation}
\eta = \frac{4mT^{1/2}}{\pi^{1/2} d} \left[ \frac{g(\nu)^{-1}}{7-3e}
+ \frac{(1+e)(3e+1)}{4(7-3e)}\nu + \left[ \frac{(1+e)(3e-1)}{8(7-3e)}
+ \frac{1}{\pi} \right](1+e)\nu^2 g(\nu) \right]~.
\end{equation}
Therefore, we find $P_{ij} = \left[ p - \xi \mathbf{\nabla}\cdot\mathbf{u} \right] \delta_{ij} - \eta D'_{ij}$.
It should be noted that Eq.(70) in Ref.\cite{jr1} should be multiplied by $rm$.
The translational energy flux is given by $q_{\alpha} = \rho a_{\alpha\beta\beta}/2 + \Theta_{\alpha\beta\beta}/2$,
where $\rho a_{\alpha\beta\beta}/2$ and $\Theta_{\alpha\beta\beta}/2$ are given by Eq.(89) and (100) in Ref.\cite{jr1}, respectively.
From Eq.(100) in Ref.\cite{jr1}, we rewrite $q_{\alpha}$ as
\begin{equation}
q_{\alpha} = \frac{1}{2}\rho a_{\alpha\beta\beta} - \xi \mathbf{\nabla}T
+ \frac{3}{2} r \nu g(\nu) \cdot \frac{1}{2}\rho a_{\alpha\beta\beta}
= \left( 1+ \frac{3}{2} r \nu g(\nu) \right) \frac{1}{2} \rho a_{\alpha\beta\beta} - \xi \mathbf{\nabla}T~.
\end{equation}
We introduce $\kappa_\rho$ and $\lambda_\rho$ as
$\frac{1}{2}\rho a_{\alpha\beta\beta} \equiv - \kappa_\rho \mathbf{\nabla}T - \lambda_\rho \mathbf{\nabla}\rho$, where
\begin{eqnarray}
\kappa_\rho &=& \frac{4mT^{1/2}}{\sigma g(\nu) r(17-15r) \pi^{1/2}} \left[ 1+\frac{3}{2} \nu g(\nu) r^2 (4r-3) \right]~, \label{kappa_rho} \\
\lambda_\rho &=& - \frac{3\sigma\pi^{1/2}(2r-1)(1-r)}{2 \nu g(\nu) (17-15r)} T^{3/2} \frac{d(\nu^2 g(\nu) )}{d\nu}~. \label{lambda_rho}
\end{eqnarray}
If we write the energy flux $q_{\alpha}$ as
$q_{\alpha} \equiv - \kappa \mathbf{\nabla}T - \lambda \mathbf{\nabla}\rho$, we obtain the heat conductivity $\kappa$
\begin{eqnarray}
\kappa = \kappa_\rho \left( 1+\frac{3}{2} r \nu g(\nu) \right) + \xi
&=& \frac{16mT^{1/2}}{\sigma \pi^{1/2}} \Big[ \frac{g(\nu)^{-1}}{(1+e)(19-15e)}
+ \frac{3(2e^2+e+1)}{8(19-15e)}\nu \nonumber \\
& & + \left\{ \frac{9(1+e)^2(2e-1)}{32(19-15e)} + \frac{1}{4\pi} \right\} (1+e) \nu^2 g(\nu) \Big]~,
\label{eq:kappa}
\end{eqnarray}
and the coefficient associated with the gradient of density $\lambda$
\begin{equation}
\lambda = \lambda_\rho \left( 1+\frac{3}{2} r \nu g(\nu) \right)
= - \frac{3\sigma \pi^{1/2} e(1-e)}{8(19-15e)}
\left[ 4g(\nu)^{-1} + 3(1+e)\nu \right] \frac{1}{\nu}\frac{d(\nu^2 g(\nu))}{d\nu} T^{3/2}~.
\end{equation}
We should note that the third term on the right hand side of Eq.(\ref{eq:kappa}) differs from our paper \cite{saitoh}.
Indeed, the coefficient $1/4\pi$ in the last term on the right hand side of Eq.~(\ref{eq:kappa}) is different from $1/2\pi$.

Now, we non-dimensionalize the static pressure, transport coefficients and
the coefficient associated with the gradient of density with the aid of $m$, $d$ and $U/2$ as
\begin{eqnarray}
p &=& \rho_p \left( \frac{U}{2} \right)^2 p^\ast(\nu) \theta, \hskip0.5em
\xi = \rho_p d \frac{U}{2} \xi^\ast(\nu) \theta^{1/2}, \hskip0.5em
\eta = \rho_p d \frac{U}{2} \eta^\ast(\nu) \theta^{1/2}, \nonumber \\
\kappa &=& \rho_p d U \kappa^\ast(\nu) \theta^{1/2}, \hskip0.5em
\lambda = d \left( \frac{U}{2} \right)^3 \lambda^\ast(\nu) \theta^{3/2}, \nonumber
\end{eqnarray}
where $p^\ast(\nu)$, $\xi^\ast(\nu)$, $\eta^\ast(\nu)$, $\kappa^\ast(\nu)$
and $\lambda^\ast(\nu)$ are dimensionless quantities listed in Table \ref{tab:subfunctions}.

\section{The Taylor expansion of the functions in Table \ref{tab:subfunctions}}
\label{ap:Taylor}

The functions in Table \ref{tab:subfunctions} are expanded into the Taylor series as
\begin{eqnarray}
g(\nu) &=& g_0 + g_1 \nu + g_2 \nu^2 + \dots~, \label{eq:Tg} \\
\theta_0 \nu^{-1} p^\ast(\nu) &=& p_0 + p_1 \nu + p_2 \nu^2 + \dots~, \label{eq:Tp} \\
\theta_0^{1/2} \nu^{-1} \xi^\ast(\nu) &=& \xi_0 + \xi_1 \nu + \xi_2 \nu^2 + \dots~, \label{eq:Txi} \\
\theta_0^{1/2} \nu^{-1} \eta^\ast(\nu) &=& \eta_0 + \eta_1 \nu + \eta_2 \nu^2 + \dots~, \label{eq:Teta} \\
\theta_0^{1/2} \nu^{-1} \kappa^\ast(\nu) &=& \kappa_0 + \kappa_1 \nu + \kappa_2 \nu^2 + \dots~, \label{eq:Tkappa} \\
\theta_0^{3/2} \nu^{-1} \lambda^\ast(\nu) &=& \lambda_0 + \lambda_1 \nu + \lambda_2 \nu^2 + \dots~. \label{eq:Tlambda}
\end{eqnarray}
Similarly, the derivatives are also expanded into the Taylor series as
\begin{eqnarray}
\theta_0 \nu^{-1} \frac{dp^\ast(\nu)}{d\nu} &=& p'_0 + p'_1 \nu + p'_2 \nu^2 + \dots~, \label{eq:Tdp} \\
\theta_0^{1/2} \nu^{-1} \frac{d\xi^\ast(\nu)}{d\nu} &=& \xi'_0 + \xi'_1 \nu + \xi'_2 \nu^2 + \dots~, \label{eq:Tdxi} \\
\theta_0^{1/2} \nu^{-1} \frac{d\eta^\ast(\nu)}{d\nu} &=& \eta'_0 + \eta'_1 \nu + \eta'_2 \nu^2 + \dots~, \label{eq:Tdeta} \\
\theta_0^{1/2} \nu^{-1} \frac{d\kappa^\ast(\nu)}{d\nu} &=& \kappa'_0 + \kappa'_1 \nu + \kappa'_2 \nu^2 + \dots~. \label{eq:Tdkappa}
\end{eqnarray}
In the following, we show the explicit expressions of the coefficients which are used in the text.
The coefficients associated with the radial distribution function are given by
\begin{equation}
g_1 = \frac{25-7\nu_0}{16(1-\nu_0)^3}~, \hskip1em
g_2 = \frac{34-7\nu_0}{16(1-\nu_0)^4}~, \hskip1em
g_3 = \frac{43-7\nu_0}{16(1-\nu_0)^5}~, \hskip1em
g_4 = \frac{52-7\nu_0}{16(1-\nu_0)^6}~. \label{eq:Tg1234}
\end{equation}
The coefficients associated with the static pressure are given by
\begin{equation}
p_1 = \frac{1}{2}(1+e)(g_0+\nu_0 g_1) \theta_0,\hskip0.1em
p_2 = \frac{1}{2}(1+e)(g_1+\nu_0 g_2) \theta_0,\hskip0.1em
p_3 = \frac{1}{2}(1+e)(g_2+\nu_0 g_3) \theta_0~.
\end{equation}
The coefficients associated with viscosity are given by
\begin{equation}
\xi_1 = \frac{1+e}{\sqrt{2\pi}}(g_0+\nu_0 g_1) \theta_0^{1/2}~, \hskip1em
\eta_1 = \left\{ a_\eta - \frac{b_\eta}{(\nu_0 g_0)^2} \right\} (g_0+\nu_0 g_1) \theta_0^{1/2}~,
\end{equation}
where
\begin{equation}
a_\eta = \sqrt{\frac{\pi}{2}}\left( \frac{(1+e)(3e-1)}{8(7-3e)} + \frac{1}{\pi} \right)(1+e)~,\hskip1em
b_\eta = \sqrt{\frac{\pi}{2}}\frac{1}{7-3e}~.
\end{equation}
The coefficients associated with the heat conductivity are given by
\begin{equation}
\kappa_1 = \left\{ a_\kappa - \frac{b_\kappa}{(\nu_0 g_0)^2} \right\} (g_0+\nu_0 g_1) \theta_0^{1/2}~,
\end{equation}
where we have introduced
\begin{equation}
a_\kappa = \sqrt{2\pi}\left( \frac{9(1+e)(2e-1)}{32(19-15e)} + \frac{1}{4\pi} \right)(1+e)~,\hskip1em
b_\kappa = \frac{\sqrt{2\pi}}{(1+e)(19-15e)}~.
\end{equation}
The coefficients associated with the derivative of the static pressure are given by
\begin{eqnarray}
p'_0 &=& \frac{1}{2\nu_0}   \left\{ (1+e)\nu_0   \left( 2 g_0 +   \nu_0 g_1 \right) + 1 \right\} \theta_0~, \label{eq:p'_0} \\
p'_1 &=& \frac{1}{2\nu_0^2} \left\{ (1+e)\nu_0^2 \left( 3 g_1 + 2 \nu_0 g_2 \right) - 1 \right\} \theta_0~, \\
p'_2 &=& \frac{1}{2\nu_0^3} \left\{ (1+e)\nu_0^3 \left( 4 g_2 + 3 \nu_0 g_3 \right) + 1 \right\} \theta_0~, \\
p'_3 &=& \frac{1}{2\nu_0^4} \left\{ (1+e)\nu_0^4 \left( 5 g_3 + 4 \nu_0 g_4 \right) - 1 \right\} \theta_0~.
\end{eqnarray}
The coefficients associated with the derivative of viscosity are given by
\begin{eqnarray}
\xi'_0 &=& \frac{1+e}{\sqrt{2\pi}}(2g_0 +  \nu_0 g_1) \theta_0^{1/2}~,\\
\eta'_0 &=& \frac{1}{\nu_0 g_0^2} \left[
- a_\eta g_1 + g_0^2 \left\{ c_\eta \nu_0 \left( 2 g_0 + \nu_0 g_1 \right) + b_\eta \right\} \right] \theta_0^{1/2}~.
\end{eqnarray}
The coefficients associated with the derivative of the heat conductivity are given by
\begin{eqnarray}
\kappa'_0 &=& \frac{1}{\nu_0 g_0^2} \left[
- a_\kappa g_1 + g_0^2 \left\{ c_\kappa \nu_0 \left( 2 g_0 + \nu_0 g_1 \right) + b_\kappa \right\} \right] \theta_0^{1/2}~, \\
\kappa'_1 &=& \frac{1}{\nu_0^2 g_0^3} \left[
a_\kappa \left\{ 2 \nu_0 g_1^2 + g_0 \left( g_1 - 2 \nu_0 g_2 \right) \right\}
+ g_0^3 \left\{ c_\kappa \nu_0^2 \left( 3g_1 + 2\nu_0 g_2 \right) - b_\kappa \right\} \right] \theta_0^{1/2}~.
\end{eqnarray}

\section{Solution of linearized equation for the non-layering mode} \label{ap:nonlayer_perturb}

In this appendix, we solve the linearized equation for the non-layering mode $(k_x\neq 0)$
\begin{equation}
\frac{\partial}{\partial t} \varphi^{\rm NL}_{\mathbf{k}(t)}=\mathcal{L}(t) \varphi^{\rm NL}_{\mathbf{k}(t)}~.
\label{ap:eq:linearized}
\end{equation}
At first, we solve the eigenvalue problem
\begin{equation}
\mathcal{L}(t) \psi^{(j)}_{\mathbf{k}(t)} = \lambda^{(j)}_{\mathbf{k}(t)} \psi^{(j)}_{\mathbf{k}(t)}~,
\label{ap:eq:nonlayer_eigen_pb}
\end{equation}
where $\lambda^{(j)}_{\mathbf{k}(t)}$ and $\psi^{(j)}_{\mathbf{k}(t)}$ $(j=1,2,3,4)$
are respectively the eigenvalues and the eigenvectors of $\mathcal{L}(t)$.
If we scale the wave number as $\mathbf{k}(t)=\epsilon\mathbf{q}(t)=\epsilon(q_x,q_y(t))$
and perturbatively solve Eq.(\ref{ap:eq:nonlayer_eigen_pb}), the eigenvalues are readily found to be
\begin{eqnarray}
\lambda^{(1)}_{\mathbf{q}(t)} &=& - \epsilon^2 r_a q(t)^2~, \\
\lambda^{(2)}_{\mathbf{q}(t)} &=& - \epsilon^2 r_b q(t)^2~, \\
\lambda^{(3)}_{\mathbf{q}(t)} &=& - \epsilon^2 r_c q(t)^2 + i \epsilon Jq(t)~, \\
\lambda^{(4)}_{\mathbf{q}(t)} &=& - \epsilon^2 r_c q(t)^2 - i \epsilon Jq(t)~,
\end{eqnarray}
which are respectively given by the eigenvectors
\begin{eqnarray}
\psi^{(1)}_{\mathbf{q}(t)} &=& \left( 0, \frac{q_y(t)}{q(t)}, -\frac{q_x}{q(t)}, 0 \right)^\mathrm{T}~, \label{eq:psi11} \\
\psi^{(2)}_{\mathbf{q}(t)} &=& \left( - \frac{p_0}{\theta_0 J}, 0, 0, \frac{p'_0}{J} \right)^\mathrm{T}~, \label{eq:psi12} \\
\psi^{(3)}_{\mathbf{q}(t)} &=& \left( \frac{\nu_0}{2J}, \frac{i}{2}\frac{q_x}{q(t)},
\frac{i}{2}\frac{q_y(t)}{q(t)}, \frac{p_0}{J} \right)^\mathrm{T}~, \label{eq:psi13} \\
\psi^{(4)}_{\mathbf{q}(t)} &=& \left( \frac{\nu_0}{2J}, -\frac{i}{2}\frac{q_x}{q(t)},
-\frac{i}{2}\frac{q_y(t)}{q(t)}, \frac{p_0}{J} \right)^\mathrm{T}~. \label{eq:psi14}
\end{eqnarray}
and the left eigenvectors
\begin{eqnarray}
\tilde{\psi}^{(1)}_{\mathbf{q}(t)} &=& \left( 0, \frac{q_y(t)}{q(t)}, -\frac{q_x}{q(t)}, 0 \right)~, \label{eq:apsi11} \\
\tilde{\psi}^{(2)}_{\mathbf{q}(t)} &=& \left( - \frac{2p_0}{J}, 0, 0, \frac{\nu_0}{J} \right)~, \label{eq:apsi12} \\
\tilde{\psi}^{(3)}_{\mathbf{q}(t)} &=& \left( \frac{p'_0}{J}, -i\frac{q_x}{q(t)},-i\frac{q_y(t)}{q(t)},
\frac{p_0}{\theta_0J} \right)~, \label{eq:apsi13} \\
\tilde{\psi}^{(4)}_{\mathbf{q}(t)} &=& \left( \frac{p'_0}{J}, i\frac{q_x}{q(t)},
i\frac{q_y(t)}{q(t)}, \frac{p_0}{\theta_0J} \right)~, \label{eq:apsi14}
\end{eqnarray}
where we defined $q(t)\equiv\sqrt{q_x^2+q_y(t)^2}$,
\begin{equation}
J \equiv \sqrt{2p_0^2/\theta_0 + \nu_0 p'_0}~, \label{ap:eq:J}
\end{equation}
and the positive constants
\begin{equation}
r_a = \frac{\eta_0}{2}~, \hskip1em
r_b = \frac{2 \nu_0p'_0\kappa_0}{J^2}~, \hskip1em
r_c = \frac{\xi_0}{2}+\frac{\eta_0}{4}+\frac{2p_0^2\kappa_0}{\theta_0J^2}~. \label{ap:eq:rarbrc}
\end{equation}

The solution of Eq.(\ref{ap:eq:linearized}) is constructed as \cite{lutsko0}
\begin{equation}
\varphi^{\rm NL}_{\mathbf{q}(t)\alpha} = \int\frac{d\mathbf{q}'(0)}{(2\pi)^2}
G_{\alpha\beta}(\mathbf{q}(t),\mathbf{q}'(0)) \varphi^{\rm NL}_{\mathbf{q}'(0)\beta}~, \label{ap:eq:formal_sol}
\end{equation}
where the indexes $\alpha, \beta$ $(=1,2,3,4)$ represent the components of $\varphi^{\rm NL}_{\mathbf{q}(t)}$
and we used the summation rule for the twice appearance of $\beta$. The Green's function is given by
\begin{equation}
G_{\alpha\beta}(\mathbf{q}(t),\mathbf{q}'(0)) = \sum_{j=1}^4 \psi_{\mathbf{q}(t)\alpha}^{(j)}
\tilde{\psi}_{\mathbf{q}'(0)\beta}^{(j)} G^{(j)}(\mathbf{q}(t),\mathbf{q}'(0)) \label{ap:eq:green}
\end{equation}
with the function $G^{(j)}(\mathbf{q}(t),\mathbf{q}'(0))$ satisfying
\begin{equation}
\left( \frac{\partial}{\partial t} + \epsilon q_x \frac{\partial}{\partial q_y(t)} \right)
G^{(j)}(\mathbf{q}(t),\mathbf{q}'(0)) = 
\lambda^{(j)}_{\mathbf{q}(t)} G^{(j)}(\mathbf{q}(t),\mathbf{q}'(0))~. \label{ap:eq:rel_Gj}
\end{equation}
Such a function $G^{(j)}(\mathbf{q}(t),\mathbf{q}'(0))$ is found to be
\begin{equation}
G^{(j)}(\mathbf{q}(t),\mathbf{q}'(0)) = (2\pi)^2 \delta(\mathbf{q}'(0)-\mathbf{q}(t))
\exp\left[ \int_0^t ds \lambda^{(j)}_{\mathbf{q}(s)} \right]~. \label{ap:eq:Gj}
\end{equation}
If we adopt $\varphi^{\rm NL}_{\mathbf{q}(0)} = \sum_{j=1}^4 \psi_{\mathbf{q}(0)}^{(j)}$ for the initial condition \cite{lutsko0},
the components of $\varphi^{\rm NL}_{\mathbf{q}(t)}$ are given by
\begin{eqnarray}
\nu_{\mathbf{q}(t)} &=& - \frac{p_0}{\theta_0J} E^{(2)}(t) + \frac{\nu_0}{J} E^{(3)}(t)\cos{\omega(t)}~,\\
u_{\mathbf{q}(t)} &=& -\frac{\epsilon t}{\sqrt{1+(\epsilon t)^2}} E^{(1)}(t) - \frac{1}{\sqrt{1+(\epsilon t)^2}} E^{(3)}(t) \sin\omega(t)~,\\
w_{\mathbf{q}(t)} &=& -\frac{1}{\sqrt{1+(\epsilon t)^2}} E^{(1)}(t) + \frac{\epsilon t}{\sqrt{1+(\epsilon t)^2}} E^{(3)}(t) \sin\omega(t)~,\\
\theta_{\mathbf{q}(t)} &=& \frac{p'_0}{J} E^{(2)}(t) + \frac{2p_0}{J} E^{(3)}(t)\cos{\omega(t)}~,
\end{eqnarray}
where we defined
\begin{eqnarray}
E^{(1)}(t) &=& \exp\left[ - q_x^2 r_a \left( \epsilon^2 t + \frac{\epsilon^4}{3} t^3 \right) \right]~, \\
E^{(2)}(t) &=& \exp\left[ - q_x^2 r_b \left( \epsilon^2 t + \frac{\epsilon^4}{3} t^3 \right) \right]~, \\
E^{(3)}(t) &=& \exp\left[ - q_x^2 r_c \left( \epsilon^2 t + \frac{\epsilon^4}{3} t^3 \right) \right]~,
\end{eqnarray}
and the frequency
\begin{equation}
\omega(t) = \frac{q_x J}{2} \left[ \epsilon t \sqrt{1+(\epsilon t)^2}
+ \ln\left\{\epsilon t + \sqrt{1+(\epsilon t)^2}\right\} \right]~.
\end{equation}

\section{Perturbative calculation of eigenvalue problem for the layering mode} \label{ap:perturb}

In this appendix, we perturbatively solve the eigenvalue problem of the layering mode
\begin{equation}
\mathcal{L}(0,k_{y0}) \varphi_{k_{y0}}^{\mathrm{L}(j)}
= \sigma^{(j)}(k_{y0}) \varphi_{k_{y0}}^{\mathrm{L}(j)}~, \label{eq:eigen_pb_ap}
\end{equation}
where $j=1,2,3,4$ and the $4 \times 4$ real matrix is given by
\begin{equation}
\mathcal{L}(0,k_{y0}) =
\begin{pmatrix}
 0 & 0 & \nu_0 k_{y0} & 0 \\
 \epsilon \frac{\eta'_0}{2} k_{y0} & - \frac{\eta_0}{2} k_{y0}^2 &
 - \epsilon & \epsilon \frac{\eta_0}{4\theta_0} k_{y0} \\
 - p'_0 k_{y0} & 0 & - ( \xi_0 + \frac{\eta_0}{2} ) k_{y0}^2 &
 - \frac{p_0}{\theta_0} k_{y0} \\
 \epsilon^2 c_1 - 2 \lambda_0 k_{y0}^2 & - 2 \epsilon \eta_0 k_{y0} &
 c_2 k_{y0} & \epsilon^2 c_3 - 2 \kappa_0 k_{y0}^2
\end{pmatrix}~.
\end{equation}
Here, $\lambda_0 \sim O(\epsilon^2)$.
The wave number is scaled as $k_{y0}=\epsilon q$ and we expand $\mathcal{L}(0,k_{y0})$,
$\sigma^{(j)}(k_{y0})$ and $\varphi_{k_{y0}}^{\mathrm{L}(j)}$ into the series of $\epsilon$ as
\begin{eqnarray}
\mathcal{L}(0,k_{y0}) &=& \epsilon \mathcal{M}_1 + \epsilon^2 \mathcal{M}_2 + \dots~, \label{eq:L_expansion_ap} \\
\sigma^{(j)}(k_{y0}) &=& \epsilon \sigma_1^{(j)} + \epsilon^2 \sigma_2^{(j)} + \dots~, \label{eq:sigma_expansion_ap} \\
\varphi_{k_{y0}}^{\mathrm{L}(j)} &=& \varphi_0^{(j)} + \epsilon \varphi_1^{(j)} + \dots~, \label{eq:varphi_expansion_ap}
\end{eqnarray}
where
\begin{equation}
\mathcal{M}_1 =
\begin{pmatrix}
 0 & 0 & \nu_0 q & 0 \\
 0 & 0 & - 1 & 0 \\
 - p'_0 q & 0 & 0 & - \frac{p_0}{\theta_0} q \\
 0 & 0 & 2 p_0 q & 0
\end{pmatrix},\hskip0.5em
\mathcal{M}_2 =
\begin{pmatrix}
 0 & 0 & 0 & 0 \\
 \frac{\eta'_0}{2} q & - \frac{\eta_0}{2} q^2 & 0 & \frac{\eta_0}{4\theta_0} q \\
 0 & 0 & - ( \xi_0 + \frac{\eta_0}{2} ) q^2 & 0 \\
 c_1 & - 2 \eta_0 q & 0 & c_3 - 2 \kappa_0 q^2
\end{pmatrix}.
\end{equation}

Substituting Eqs.(\ref{eq:L_expansion_ap})-(\ref{eq:sigma_expansion_ap}) into Eq.(\ref{eq:eigen_pb_ap}),
we find the first nonzero terms
\begin{equation}
\mathcal{M}_1 \varphi_0^{(j)} = \sigma_1^{(j)} \varphi_0^{(j)}~,
\label{eq:O1_ap}
\end{equation}
at $O(\epsilon)$. From Eq.(\ref{eq:O1_ap}), we find the eigenvalues
\begin{equation}
\sigma_1^{(1)} = \sigma_1^{(2)}=0~, \hskip2em
\sigma_1^{(3)} = - \sigma_1^{(4)} = iJq~.
\label{eq:sigma1_ap}
\end{equation}
The eigenvalues Eq.(\ref{eq:sigma1_ap}) are given by the eigenvectors
\begin{eqnarray}
\varphi_0^{(1)} &=& \left( 0, 1, 0, 0 \right)^\mathrm{T}~, \label{eq:varphi01} \\
\varphi_0^{(2)} &=& \left( - \frac{p_0}{\theta_0J}, 0, 0, \frac{p'_0}{J} \right)^\mathrm{T}~, \label{eq:varphi02} \\
\varphi_0^{(3)} &=& \left( \frac{\nu_0}{2J}, -\frac{1}{2 J q}, \frac{i}{2}, \frac{p_0}{J} \right)^\mathrm{T}~, \label{eq:varphi03} \\
\varphi_0^{(4)} &=& \left( \frac{\nu_0}{2J}, -\frac{1}{2 J q}, -\frac{i}{2}, \frac{p_0}{J} \right)^\mathrm{T}~, \label{eq:varphi04}
\end{eqnarray}
respectively, and the corresponding left eigenvectors are given by
\begin{eqnarray}
\tilde{\varphi}_0^{(1)} &=& \left( \frac{p'_0}{J^2 q}, 1, 0, \frac{p_0}{\theta_0 J^2 q} \right)~, \label{eq:avarphi01} \\
\tilde{\varphi}_0^{(2)} &=& \left( - \frac{2 p_0}{J}, 0, 0, \frac{\nu_0}{J} \right)~, \label{eq:avarphi02} \\
\tilde{\varphi}_0^{(3)} &=& \left( \frac{p'_0}{J}, 0, -i, \frac{p_0}{\theta_0J} \right)~, \label{eq:avarphi03} \\
\tilde{\varphi}_0^{(4)} &=& \left( \frac{p'_0}{J}, 0, i, \frac{p_0}{\theta_0J} \right)~, \label{eq:avarphi04}
\end{eqnarray}
respectively.
The eigenvectors Eqs.(\ref{eq:varphi01})-(\ref{eq:avarphi04}) are orthogonal and normalized, i.e.
$\tilde{\varphi}_0^{(j)} \varphi_0^{(k)} = \delta_{jk}$, where $j,k=1,2,3,4$ and $\delta_{jk}$ is the Kronecker delta.
Because the critical eigenvalue is a real number, we are interested in the case of $\sigma_1^{(1)} = \sigma_1^{(2)}=0$.
However, $\varphi_0^{(1)}$ and $\varphi_0^{(2)}$ are degenerated, thus we rewrite Eq.(\ref{eq:varphi_expansion_ap}) as
\begin{equation}
\varphi_{k_{y0}}^{\mathrm{L}(l)} = \left\{ a_1^{(l)} \varphi_0^{(1)} + a_2^{(l)} \varphi_0^{(2)} \right\} + \epsilon \varphi_1^{(l)} + \dots~,
\label{eq:varphi_expansion2_ap}
\end{equation}
where $l=1,2$ and the coefficients $a_1^{(l)}$ and $a_2^{(l)}$ are determined later.

At $O(\epsilon^2)$ of Eq.(\ref{eq:eigen_pb_ap}), we find
\begin{equation}
\mathcal{M}_1 \varphi_1^{(l)} + \mathcal{M}_2 \left\{ a_1^{(l)} \varphi_0^{(1)} + a_2^{(l)} \varphi_0^{(2)} \right\}
= \sigma_2^{(l)} \left\{ a_1^{(l)} \varphi_0^{(1)} + a_2^{(l)} \varphi_0^{(2)} \right\}~.
\label{eq:O2_ap}
\end{equation}
If we respectively multiply $\tilde{\varphi}_0^{(l)}$~$(l=1,2)$ to Eq.(\ref{eq:O2_ap}), we find
\begin{equation}
\mathbf{M}_2 \mathbf{a}^{(l)} = \sigma_2^{(l)} \mathbf{a}^{(l)}~,
\label{eq:O2_matrix_ap}
\end{equation}
where $\mathbf{a}^{(l)} \equiv (a_1^{(l)},a_2^{(l)})^\mathrm{T}$ and
\begin{equation}
\mathbf{M}_2 =
\begin{pmatrix}
\tilde{\varphi}_0^{(1)} \mathcal{M}_2 \varphi_0^{(1)} & \tilde{\varphi}_0^{(1)} \mathcal{M}_2 \varphi_0^{(2)} \\
\tilde{\varphi}_0^{(2)} \mathcal{M}_2 \varphi_0^{(1)} & \tilde{\varphi}_0^{(2)} \mathcal{M}_2 \varphi_0^{(2)}
\end{pmatrix}~.
\label{eq:M2_ap}
\end{equation}
From Eq.(\ref{eq:O2_matrix_ap}), we find the eigenvalues
\begin{equation}
\sigma_2^{(1)} = - \frac{1}{4\theta_0J^2} \left(f_1 - \sqrt{f_1^2 - f_2} \right), \hskip0.5em
\sigma_2^{(2)} = - \frac{1}{4\theta_0J^2} \left(f_1 + \sqrt{f_1^2 - f_2} \right),
\label{eq:sigma2_ap}
\end{equation}
which are given by the eigenvectors
\begin{equation}
\mathbf{a}^{(1)} = c_a \left( f_3 - \sqrt{f_1^2 - f_2}, 8 \nu_0 \theta_0^{1/2} \eta_0 J q \right)^\mathrm{T}, \hskip0.5em
\mathbf{a}^{(2)} = c_a \left( f_3 + \sqrt{f_1^2 - f_2}, 8 \nu_0 \theta_0^{1/2} \eta_0 J q \right)^\mathrm{T},
\label{eq:a12_ap}
\end{equation}
respectively, where $f_1$, $f_2$, $f_3$ and $c_a$ are listed in Table \ref{tab:subfunctions_ap}.

Because $\sigma_2^{(1)} - \sigma_2^{(2)} = \sqrt{f_1^2 - f_2}/2\theta_0J^2 >0$, the growth rate is given by $\sigma_2^{(1)}$.
If we expand $\sigma_2^{(1)}$ into the series of the scaled wave number $q$, we find the dispersion relation
\begin{equation}
\sigma_2^{(1)} = r_2 q^2 + r_4 q^4 + O(q^6)~,
\end{equation}
where the coefficients $r_2$ and $r_4$ are respectively given by
\begin{eqnarray}
r_2 &=& \frac{\nu_0 \eta_0 ( 2\eta'_0 p_0 - \eta_0 p'_0 - p_0 c_1 + \theta_0 p'_0 c_3)}
{2 \nu_0 (p_0 c_1 - \theta_0 p'_0 c_3) + 4 p_0 \eta_0}~, \label{ap:eq:r2}\\
r_4 &=& - \frac{\nu_0\theta_0\eta_0\{ \nu_0\eta_0p'_0 - 2p_0(\eta_0 + \nu_0\eta'_0) \}}{4\{\nu_0(p_0c_1-\theta_0p'_0c_3) + 2p_0\eta_0\}^3}
\Big[ (c_1-2\eta'_0)p_0\eta_0J^2 + 4c_3\nu_0\theta_0{p'_0}^2\kappa_0 \\
& & + p'_0 \{ \eta_0^2 J^2-4c_1\nu_0p_0\kappa_0-\eta_0 (\theta_0c_3J^2+8p_0\kappa_0)\} \Big]~. \label{ap:eq:r4}
\end{eqnarray}

If we multiply $\varphi_0^{(n)}\tilde{\varphi}_0^{(n)}$ $(n=3,4)$ to Eq.(\ref{eq:O2_ap}), we find $\varphi_1^{(1)}$ as
\begin{equation}
\varphi_1^{(1)} = - \sum_{n=3,4} \frac{1}{\sigma_1^{(n)}} \varphi_0^{(n)}
\left[ \tilde{\varphi}_0^{(n)} \mathcal{M}_2 \left\{ a_1^{(1)} \varphi_0^{(1)} + a_2^{(1)} \varphi_0^{(2)} \right\} \right]
\equiv C_{\varphi_1} (0,0,1,0)^\mathrm{T}~,
\end{equation}
where
\begin{equation}
C_{\varphi_1} = \frac{2p_0\eta_0}{\theta_0J^2} a_1^{(1)} + \frac{(2\theta_0p'_0\kappa_0q^2 + p_0c_1
- \theta_0p'_0c_3)p_0}{\theta_0^{3/2}J^3q}a_2^{(1)}~. \label{eq:C_varphi1_ap}
\end{equation}
Therefore, the eigenvector Eq.(\ref{eq:varphi_expansion_ap}) truncated at $O(\epsilon)$ is given by
\begin{equation}
\varphi_{k_{y0}}^{\mathrm{L}(1)} =(\nu_q, u_q, w_q, \theta_q)^\mathrm{T}
= \left( - \frac{p_0}{\theta_0J} a_2^{(1)}, a_1^{(1)}, \epsilon C_{\varphi_1}, \frac{p'_0}{J} a_2^{(1)} \right)^\mathrm{T}~.
\label{eq:varphiq_ap}
\end{equation}

In the same way, we calculate the left eigenvector $\tilde{\varphi}_{k_{y0}}^{(l)}$ of Eq.(\ref{eq:eigen_pb_ap}).
Because $\tilde{\varphi}_0^{(1)}$ and $\tilde{\varphi}_0^{(2)}$ are also degenerated
to the same eigenvalue $\sigma_1^{(1)} = \sigma_1^{(2)}=0$, we write the left eigenvector as
\begin{equation}
\tilde{\varphi}_{k_{y0}}^{\mathrm{L}(l)} = \left\{ \tilde{a}_1^{(l)} \tilde{\varphi}_0^{(1)}
+ \tilde{a}_2^{(l)} \tilde{\varphi}_0^{(2)} \right\} + \epsilon \tilde{\varphi}_1^{(l)} + \dots~,
\label{eq:avarphi_expansion2_ap}
\end{equation}
where $l=1,2$ and the coefficients $\tilde{a}_1^{(l)}$ and $\tilde{a}_2^{(l)}$ are determined later.
At $O(\epsilon^2)$, we find
\begin{equation}
\tilde{\varphi}_1^{(l)} \mathcal{M}_1
+ \left\{ \tilde{a}_1^{(l)} \tilde{\varphi}_0^{(1)} + \tilde{a}_2^{(l)} \tilde{\varphi}_0^{(2)} \right\} \mathcal{M}_2
= \sigma_2^{(l)} \left\{ \tilde{a}_1^{(l)} \tilde{\varphi}_0^{(1)} + \tilde{a}_2^{(l)} \tilde{\varphi}_0^{(2)} \right\}~.
\label{eq:O2_adjoint_ap}
\end{equation}
If we respectively multiply $\varphi_0^{(l)}$~$(l=1,2)$ to Eq.(\ref{eq:O2_adjoint_ap}), we find
\begin{equation}
\mathbf{M}_2^\mathrm{T} \tilde{\mathbf{a}}^{(l)} = \sigma_2^{(l)} \tilde{\mathbf{a}}^{(l)}~,
\label{eq:O2_adjoint_matrix_ap}
\end{equation}
where $\tilde{\mathbf{a}}^{(l)} \equiv (\tilde{a}_1^{(l)},\tilde{a}_2^{(l)})^\mathrm{T}$.
Then, we solve the eigenvalue problem Eq.(\ref{eq:O2_adjoint_matrix_ap}) and find
\begin{equation}
\tilde{\mathbf{a}}^{(1)} = \tilde{c}_a \left( f_3 - \sqrt{f_1^2 - f_2}, \frac{f_4}{\theta_0^{1/2}Jq} \right)^\mathrm{T}, \hskip0.5em
\tilde{\mathbf{a}}^{(2)} = \tilde{c}_a \left( f_3 + \sqrt{f_1^2 - f_2}, \frac{f_4}{\theta_0^{1/2}Jq} \right)^\mathrm{T},
\label{eq:b12_ap}
\end{equation}
where $f_4$ and $\tilde{c}_a$ are listed in Table \ref{tab:subfunctions_ap}.
If we multiply $\varphi_0^{(n)}\tilde{\varphi}_0^{(n)}$ $(n=3,4)$ to Eq.(\ref{eq:O2_adjoint_ap}),
we find $\tilde{\varphi}_1^{(1)}$ as
\begin{equation}
\tilde{\varphi}_1^{(1)} = - \sum_{n=3,4} \frac{1}{\sigma_1^{(n)}}
\left[ \left\{ \tilde{a}_1^{(1)} \tilde{\varphi}_0^{(1)}
+ \tilde{a}_2^{(1)} \tilde{\varphi}_0^{(2)} \right\} \mathcal{M}_2 \varphi_0^{(n)} \right] \tilde{\varphi}_0^{(n)}
\equiv \tilde{C}_{\varphi_1} (0,0,1,0)~,
\end{equation}
where
\begin{equation}
\tilde{C}_{\varphi_1} = \frac{f_5}{2\theta_0J^4q^2} \tilde{a}_1^{(1)}
- \frac{\nu_0\{ 4p_0\kappa_0q^2 - 2\eta_0-\nu_0c_1-2p_0c_3 \}}{\theta_0^{1/2}J^3q}\tilde{a}_2^{(1)}
\label{eq:aC_varphi1_ap}
\end{equation}
and $f_5$ is given in Tab.\ref{tab:subfunctions_ap}.
Therefore, the left eigenvector of Eq.(\ref{eq:eigen_pb_ap}) truncated at $O(\epsilon^2)$ is given by
\begin{equation}
\tilde{\varphi}_{k_{y0}}^{\mathrm{L}(1)} = (\tilde{\nu}_q, \tilde{u}_q, \tilde{w}_q, \tilde{\theta}_q)
=\left( \frac{p'_0}{J^2 q}\tilde{a}_1^{(1)} - \frac{2p_0}{J}\tilde{a}_2^{(1)},
\tilde{a}_1^{(1)}, \epsilon \tilde{C}_{\varphi_1}, \frac{p_0}{\theta_0J^2 q}\tilde{a}_1^{(1)}+\frac{\nu_0}{J}\tilde{a}_2^{(1)} \right)~.
\label{eq:adjointq_ap}
\end{equation}

\begin{figure*}
\begin{center}
\includegraphics[width=\textwidth]{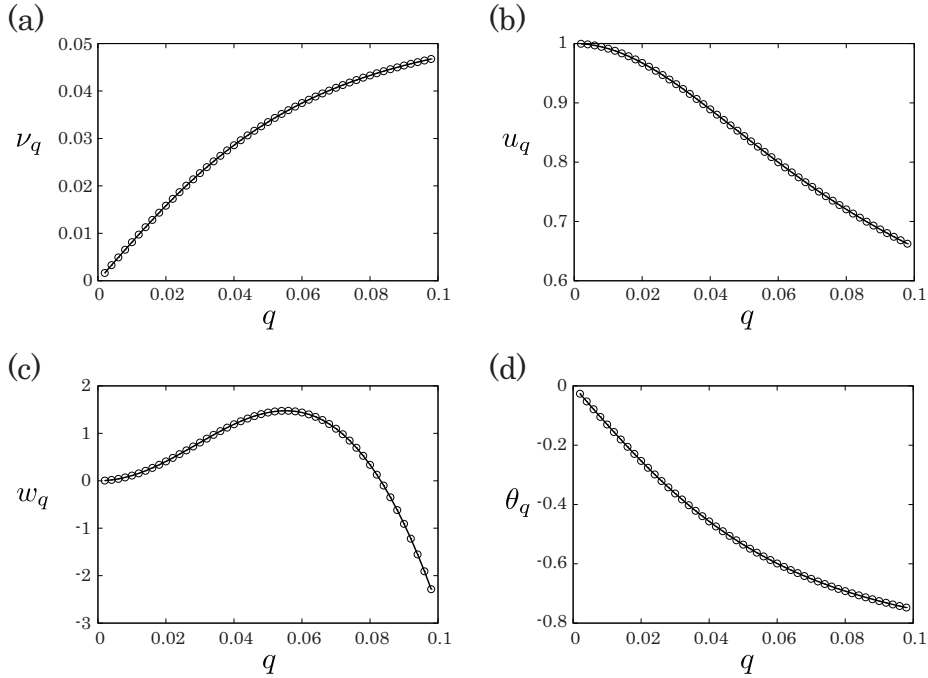}
\caption{The components of the eigenvector (a)$\nu_q$, (b)$u_q$, (c)$w_q$ and (d)$\theta_q$
as the functions of $q$, where the open circles and solid lines represent the numerical results
and analytic form Eq.(\ref{eq:varphiq_ap}), respectively. Here, we used $\epsilon=0.01$, $e=0.99$ and $\nu_0=0.4$.}
\label{fig:egvectors_ap}
\end{center}
\end{figure*}

Let us compare the analytic form Eq.(\ref{eq:varphiq_ap}) with the numerical result. Figure \ref{fig:egvectors_ap}
shows the components of the eigenvector (a)$\nu_q$, (b)$u_q$, (c)$w_q$ and (d)$\theta_q$ as the functions
of $q$, where the open circles and solid lines represent the numerical results and analytic forms, respectively.
From these results, Eq.(\ref{eq:varphiq_ap}) well describes the numerical results. We also confirmed that
Eq.(\ref{eq:adjointq_ap}) well reproduces the results of the numerical calculations.
\begin{table}
\caption{The functions $f_1$, $f_2$, $f_3$, $f_4$, $f_5$, $f_6$, $c_a$ and $\tilde{c}_a$ in the text.
Here, $c_a$ and $\tilde{c}_a$ are determined by the normalized condition of the eigenvector.}
\label{tab:subfunctions_ap}
\begin{tabular}{lll}
\hline\noalign{\smallskip}
$f_1$ &$=$& $\theta_0(\eta_0 J^2 + 4\nu_0 p'_0 \kappa_0)q^2 + 4\eta_0 p_0 + 2\nu_0 p_0 c_1 - 2\nu_0\theta_0 p'_0 c_3$ \\
$f_2$ &$=$& $16\nu_0\theta_0^2\eta_0 p'_0 \kappa_0 J^2 q^4 + 8\nu_0\theta_0\eta_0 J^2 \{ p_0(c_1 - 2\eta'_0)+p'_0(\eta_0-\theta_0 c_3) \} q^2$ \\
$f_3$ &$=$& $\theta_0(\eta_0 J^2 - 4\nu_0p'_0\kappa_0)q^2 + 4\eta_0p_0 - 2\nu_0p_0c_1 + 2\nu_0\theta_0p'_0c_3$ \\
$f_4$ &$=$& $\theta_0\{ 8p_0p'_0\kappa_0 + (2p_0\eta'_0-p'_0\eta_0)J^2 \}q^2 + 4p_0(p_0c_1-\theta_0p'_0c_3)$ \\
$f_5$ &$=$& $\{ \nu_0\theta_0\eta'_0J^2-8p_0^2\kappa_0+\eta_0J^2(p_0+\theta_0) \}q^2+ 2p_0(2\eta_0+\nu_0c_1+2p_0c_3)$ \\
$f_6$ &$=$& $f_3 - (f_1^2 - f_2)^{1/2}$ \\
$c_a$ &$=$& $- \left\{ 64\nu_0^2\eta_0^2q^2 (p_0^2 + \theta_0^2 {p'_0}^2 ) + f_6^2 \right\}^{-1/2}$ \\
$\tilde{c}_a$ &$=$& $- \theta_0 J^2 q \left\{(\nu_0^2+4p_0^2)f_4^2
+ 2(\nu_0p_0-2\theta_0p_0p'_0)f_4f_6 + (\theta_0^2J^4q^2+\theta_0^2{p'_0}^2+p_0^2) f_6^2 \right\}^{-1/2}$ \\
\noalign{\smallskip}\hline
\end{tabular}
\end{table}
\bibliographystyle{spphys}       

\begin{thebibliography}{10}
\providecommand{\url}[1]{{#1}}
\providecommand{\urlprefix}{URL }
\expandafter\ifx\csname urlstyle\endcsname\relax
  \providecommand{\doi}[1]{DOI \discretionary{}{}{}#1}\else
  \providecommand{\doi}{DOI \discretionary{}{}{}\begingroup
  \urlstyle{rm}\Url}\fi

\bibitem{luding1}
S.~Luding, Nonlinearity \textbf{22}, R101 (2009)

\bibitem{luding2}
T.~P{\"o}schel, S.~Luding (eds.), \emph{Granular Gases} (Springer-Verlag,
  Berlin, 2001)

\bibitem{bri}
N.V. Brilliantov, T.~P{\"o}schel, \emph{Kinetic Theory of Granular Gases}
  (Oxford University Press, Oxford, 2004)

\bibitem{gold}
I.~Goldhirsch, Annu.\ Rev.\ Fluid Mech. \textbf{35}, 267 (2003)

\bibitem{jeager}
H.~Jeager, S.~Nagel, R.~Behringer, Rev.\ Mod.\ Phys. \textbf{68}, 1259 (1996)

\bibitem{chong}
S.~Chong, M.~Otsuki, H.~Hayakawa, S.~Luding, Phys.\ Rev.\ E \textbf{81}, 041130
  (2010)

\bibitem{fort}
Y.~Forterre, O.~Pouliquen, Annu.\ Rev.\ Fluid Mech. \textbf{40}, 1 (2008)

\bibitem{poul}
O.~Pouliquen, Phys.\ Fluids \textbf{11}, 542 (1999)

\bibitem{sela}
N.~Sela, I.~Goldhirsch, S.H. Noskowicz, Phys.\ Fluids \textbf{8}, 2337 (1996)

\bibitem{santos}
A.~Santos, V.~Garz{\'o}, J.W. Dufty, Phys.\ Rev.\ E \textbf{69}, 061303 (2004)

\bibitem{tan}
M.L. Tan, I.~Goldhirsch, Phys.\ Fluids \textbf{9}, 856 (1997)

\bibitem{saitoh}
K.~Saitoh, H.~Hayakawa, Phys.\ Rev.\ E \textbf{75}, 021302 (2007)

\bibitem{kumaran1}
V.~Kumaran, Phys.\ Rev.\ Lett. \textbf{96}, 258002 (2006)

\bibitem{kumaran2}
V.~Kumaran, Phys.\ Rev.\ E \textbf{79}, 011301 (2009)

\bibitem{kumaran3}
V.~Kumaran, Phys.\ Rev.\ E \textbf{79}, 011302 (2009)

\bibitem{orpe1}
A.~Orpe, A.~Kudrolli, Phys.\ Rev.\ Lett. \textbf{98}, 238001 (2007)

\bibitem{orpe2}
A.~Orpe, V.~Kumaran, K.~Reddy, A.~Kudrolli, Europhys.\ Lett. \textbf{84}, 64003
  (2008)

\bibitem{rycroft}
C.~Rycroft, A.~Orpe, A.~Kudrolli, Phys.\ Rev.\ E \textbf{80}, 031305 (2009)

\bibitem{lutsko0}
J.F. Lutsko, J.W. Dufty, Phys.\ Rev.\ A \textbf{32}, 3040 (1985)

\bibitem{otsuki0}
M.~Otsuki, H.~Hayakawa, Eur.\ Phys.\ J.\ Special Topics \textbf{179}, 179
  (2009)

\bibitem{otsuki1}
M.~Otsuki, H.~Hayakawa, Phys.\ Rev.\ E \textbf{79}, 021502 (2009)

\bibitem{otsuki2}
M.~Otsuki, H.~Hayakawa, J.\ Stat.\ Mech: Theor.\ Exp. p. L08003 (2009)

\bibitem{louge1}
M.Y. Louge, Phys.\ Fluids \textbf{6}, 2253 (1994)

\bibitem{louge2}
M.Y. Louge, Phys.\ Rev.\ E \textbf{67}, 061303 (2003)

\bibitem{louge3}
H.~Xu, A.P. Reeves, M.Y. Louge, Rev.\ Sci.\ Instrum. \textbf{75}, 811 (2004)

\bibitem{louge4}
H.~Xu, M.Y. Louge, A.P. Reeves, Continuum Mech.\ Thermodyn. \textbf{15}, 321
  (2003)

\bibitem{khain1}
E.~Khain, Phys.\ Rev.\ E \textbf{75}, 051310 (2007)

\bibitem{khain2}
E.~Khain, Eur.\ Phys.\ Lett. \textbf{87}, 14001 (2009)

\bibitem{midi}
G.~Midi, Eur.\ Phys.\ J.\ E \textbf{14}, 341 (2004)

\bibitem{cruz}
F.~da~Cruz, S.~Eman, M.~Prochnow, J.~Roux, F.~Chevoir, Phys.\ Rev.\ E
  \textbf{72}, 021309 (2005)

\bibitem{hatano1}
T.~Hatano, Phys.\ Rev.\ E \textbf{75}, 060301(R) (2007)

\bibitem{hecke}
M.~van Hecke, J.\ Phys.\: Condens.\ Matter \textbf{22}, 033101 (2010)

\bibitem{hatano2}
T.~Hatano, M.~Otsuki, S.~Sasa, J.\ Phys.\ Soc.\ Jpn. \textbf{76}, 023001 (2007)

\bibitem{hatano3}
T.~Hatano, J.\ Phys.\ Soc.\ Jpn. \textbf{77}, 123002 (2008)

\bibitem{otsuki3}
M.~Otsuki, H.~Hayakawa, Prog.\ Theor.\ Phys. \textbf{121}, 647 (2009)

\bibitem{otsuki4}
M.~Otsuki, H.~Hayakawa, Phys.\ Rev.\ E \textbf{80}, 011308 (2009)

\bibitem{otsuki5}
M.~Otsuki, H.~Hayakawa, S.~Luding, Prog.\ Theor.\ Phys.\ Suppl. \textbf{184},
  110 (2010)

\bibitem{lun}
C.K.K. Lun, J.\ Fluid Mech. \textbf{233}, 539 (1991)

\bibitem{dufty1}
J.J. Brey, J.W. Dufty, C.S. Kim, A.~Santos, Phys.\ Rev.\ E \textbf{58}, 4638
  (1998)

\bibitem{dufty2}
V.~Garz{\'o}, J.W. Dufty, Phys.\ Rev.\ E \textbf{59}, 5895 (1998)

\bibitem{lutsko1}
J.F. Lutsko, Phys.\ Rev.\ E \textbf{70}, 061101 (2004)

\bibitem{lutsko2}
J.F. Lutsko, Phys.\ Rev.\ E \textbf{72}, 021306 (2005)

\bibitem{lutsko3}
J.F. Lutsko, Phys.\ Rev.\ E \textbf{73}, 021302 (2006)

\bibitem{jr1}
J.T. Jenkins, M.W. Richman, Phys.\ Fluids \textbf{28}, 3485 (1985)

\bibitem{jr2}
J.T. Jenkins, M.W. Richman, Arch.\ Ration.\ Mech.\ Anal. \textbf{87}, 355
  (1985)

\bibitem{linear1}
S.B. Savage, J.\ Fluid Mech. \textbf{241}, 109 (1992)

\bibitem{linear2}
V.~Garz{\'o}, Phys.\ Rev.\ E \textbf{73}, 021304 (2006)

\bibitem{layer1}
P.J. Schmid, H.K. Kyt{\"o}maa, J.\ Fluid Mech. \textbf{264}, 255 (1994)

\bibitem{layer2}
C.H. Wang, R.~Jackson, S.~Sundaresan, J.\ Fluid Mech. \textbf{308}, 31 (1996)

\bibitem{layer3}
M.~Alam, P.R. Nott, J.\ Fluid Mech. \textbf{343}, 267 (1997)

\bibitem{layer4}
M.~Alam, P.R. Nott, J.\ Fluid Mech. \textbf{377}, 99 (1998)

\bibitem{layer5}
B.~Gayen, M.~Alam, J.\ Fluid Mech. \textbf{567}, 195 (2006)

\bibitem{otsuki6}
M.~Otsuki, H.~Hayakawa, Phys.\ Rev.\ E \textbf{83}, 051301 (2011)

\bibitem{shukla1}
P.~Shukla, M.~Alam, Phys.\ Rev.\ Lett. \textbf{103}, 068001 (2009)

\bibitem{shukla2}
P.~Shukla, M.~Alam, J.\ Fluid Mech. \textbf{666}, 204 (2011)

\bibitem{shukla3}
P.~Shukla, M.~Alam, J.\ Fluid Mech. \textbf{672}, 147 (2011)

\bibitem{reynolds}
W.C. Reynolds, M.C. Potter, J.\ Fluid Mech. \textbf{27}, 465 (1967)

\bibitem{lees}
A.W. Lees, S.F. Edwards, J.\ Phys.\ C \textbf{5}, 1921 (1972)

\bibitem{stuart}
J.T. Stuart, J.\ Fluid Mech. \textbf{9}, 353 (1960)

\bibitem{ss}
K.~Stewartson, J.T. Stuart, J.\ Fluid Mech. \textbf{48}, 529 (1971)

\bibitem{nw}
A.C. Newell, J.A. Whitehead, J.\ Fluid Mech. \textbf{38}, 279 (1969)

\bibitem{reduction1}
Y.~Kuramoto, \emph{Chemical Oscillations, Waves and Turbulence} (Springer,
  Berlin, 1984)

\bibitem{reduction2}
I.S. Aranson, L.~Kramer, Rev.\ Mod.\ Phys. \textbf{74}, 99 (2002)

\bibitem{reduction3}
M.~Cross, P.~Hohenberg, Rev.\ Mod.\ Phys. \textbf{65}, 851 (1993)

\bibitem{jz}
J.~Jenkins, C.~Zhang, Phys.\ Fluids \textbf{14}, 1228 (2002)

\bibitem{yj}
D.~Yoon, J.~Jenkins, Phys.\ Fluids \textbf{17}, 083301 (2005)

\bibitem{gnu4}
L.~Verlet, D.~Levesque, Mol.\ Phys. \textbf{46}, 969 (1982)

\bibitem{gnu3}
D.~Henderson, Mol.\ Phys. \textbf{34}, 301 (1977)

\bibitem{gnu2}
D.~Henderson, Mol.\ Phys. \textbf{30}, 971 (1975)

\bibitem{gnu1}
N.~Carnahan, K.~Starling, J.\ Chem.\ Phys. \textbf{51}, 635 (1969)

\bibitem{lapack}
E.~Anderson, Z.~Bai, C.~Bischof, S.~Blackford, J.~Demmel, J.~Dongarra, J.D.
  Croz, A.~Greenbaum, S.~Hammarling, A.~McKenney, D.~Sorensen, \emph{{LAPACK}
  Users' Guide}, 3rd edn. (Society for Industrial and Applied Mathematics,
  Philadelphia, PA, 1999)

\bibitem{genamp1}
M.C. Cross, P.G. Daniels, P.C. Hohenberg, E.D. Siggia, J.\ Fluid Mech.
  \textbf{127}, 155 (1983)

\bibitem{genamp2}
W.~van Saarloos, Phys.\ Rev.\ A \textbf{39}, 6367 (1989)

\bibitem{genamp3}
T.S. Komatsu, H.~Hayakawa, Phys.\ Lett.\ A \textbf{183}, 56 (1993)

\end{thebibliography}
%
%


\end{document}